\documentclass[11pt,a4paper]{article}
\pdfoutput=1

\usepackage[utf8]{inputenc}

\usepackage{jheppub}

\usepackage{graphicx,amssymb,amsmath,mathtools,bm,bbm}
\usepackage[labelformat=simple]{subcaption}
\usepackage{autobreak}
\usepackage{hepunits}
\usepackage{physics}
\usepackage{epsfig}

\usepackage{hyperref}
\usepackage{url}
\hypersetup{pdfnewwindow=true}
\usepackage[utf8]{inputenc}

\def\be{\begin{equation}}
\def\ee{\end{equation}}

\allowdisplaybreaks

\preprint{
  \begin{flushright}
    P3H-20-01 \\
    TTK-20-01
  \end{flushright}}

\title{Exact quark-mass dependence of the Higgs-gluon form factor at
  three loops in QCD}

\author{Micha\l\ Czakon}
\author{and Marco Niggetiedt}
\emailAdd{mczakon@physik.rwth-aachen.de}
\emailAdd{marco.niggetiedt@rwth-aachen.de}
\affiliation{Institut f\"{u}r Theoretische Teilchenphysik und
  Kosmologie,
  RWTH Aachen University, \\
  D-52056 Aachen, Germany}

\abstract{We determine the three-loop form factor parameterising the
  amplitude for the production of an off-shell Higgs boson in gluon
  fusion in QCD with a single massive quark. The result is obtained
  via a numerical solution of a system of differential
  equation for the occurring master integrals. The solution is also
  used to determine the high-energy and threshold expansions of the
  form factor. Our findings may be used for the evaluation of virtual
  corrections generated by top-quark and b-quark loops in Higgs boson
  hadroproduction cross sections at next-to-next-to-leading order.}

\begin{document}

\maketitle

\newpage


\section{Introduction}

Recent interest in the Higgs-gluon form factor is stimulated primarily
by studies on the precision of cross section predictions for various
hadron-collider processes involving an intermediate Higgs boson
\cite{deFlorian:2016spz}. Indeed, the amplitude $gg \to H$ contributes
to both single- and double-Higgs production with subsequent Higgs
decay to a pair of fermions or off-shell gauge bosons. In consequence,
applications require the knowledge of the form factor for arbitrary
virtualities, and the uncertainty induced by the standard use of the
infinite top-quark mass limit plays a non-negligible role.

\begin{figure}[t]
    \centering
    \begin{subfigure}[b]{.35\textwidth}
        \centering
        \includegraphics[scale=.8]{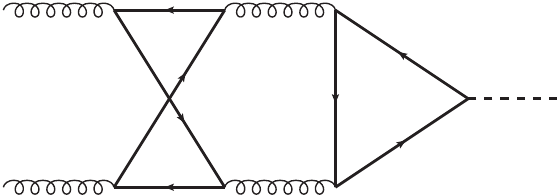}
    \end{subfigure}%
        \begin{subfigure}[b]{.35\textwidth}
        \centering
        \includegraphics[scale=.8]{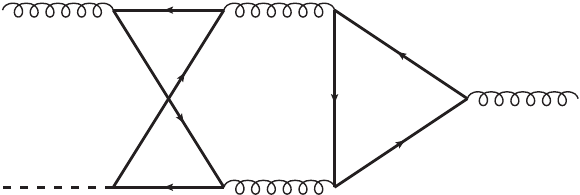}
    \end{subfigure}%
        \begin{subfigure}[b]{.35\textwidth}
        \centering
        \includegraphics[scale=.8]{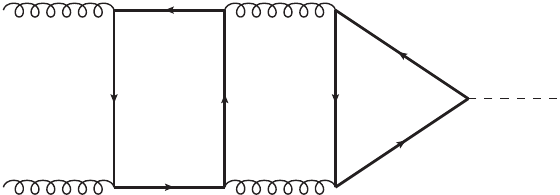}
    \end{subfigure}%
    \\[14pt]
    \centering
    \begin{subfigure}[b]{.35\textwidth}
        \centering
        \includegraphics[scale=.8]{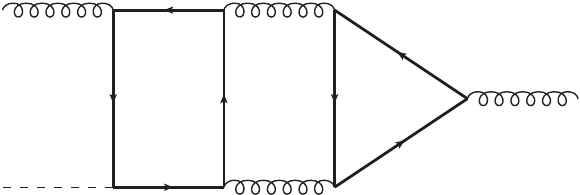}
    \end{subfigure}%
        \begin{subfigure}[b]{.35\textwidth}
        \centering
        \includegraphics[scale=.8]{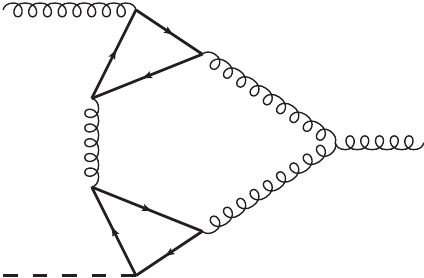}
    \end{subfigure}%
        \begin{subfigure}[b]{.35\textwidth}
        \centering
        \includegraphics[scale=.8]{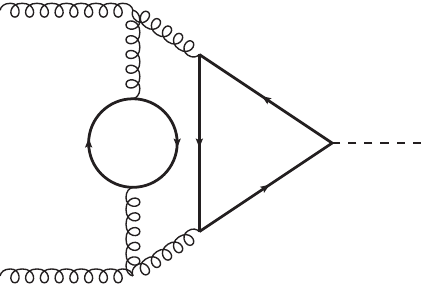}
    \end{subfigure}%
    \\[14pt]
    \centering
    \begin{subfigure}[b]{.35\textwidth}
        \centering
        \includegraphics[scale=.8]{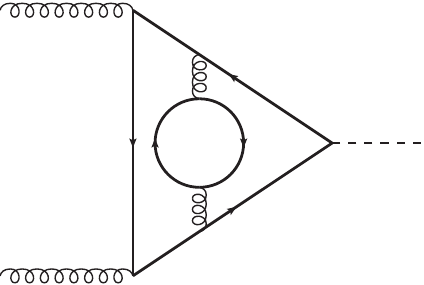}
    \end{subfigure}%
        \begin{subfigure}[b]{.35\textwidth}
        \centering
        \includegraphics[scale=.8]{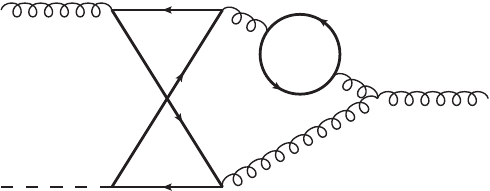}
    \end{subfigure}%
        \begin{subfigure}[b]{.35\textwidth}
        \centering
        \includegraphics[scale=.8]{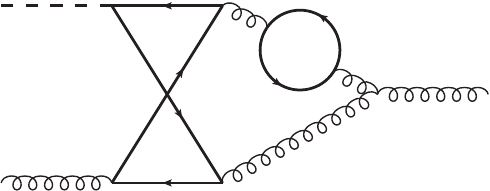}
    \end{subfigure}%
    \\[14pt]
    \centering
    \begin{subfigure}[b]{.35\textwidth}
        \centering
        \includegraphics[scale=.8]{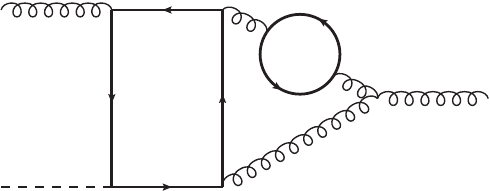}
    \end{subfigure}%
        \begin{subfigure}[b]{.35\textwidth}
        \centering
        \includegraphics[scale=.8]{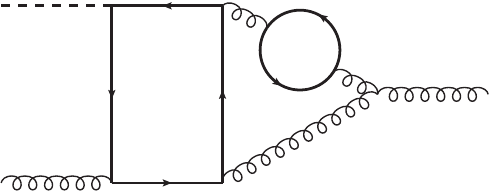}
    \end{subfigure}%
        \begin{subfigure}[b]{.35\textwidth}
        \centering
        \includegraphics[scale=.8]{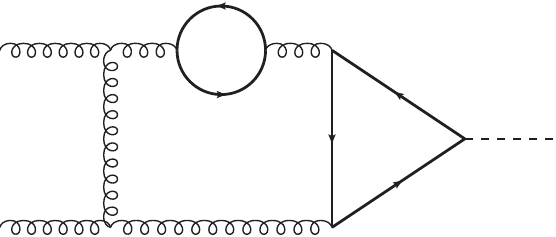}
    \end{subfigure}%
    \\[14pt]
    \centering
    \begin{subfigure}[b]{.35\textwidth}
        \centering
        \includegraphics[scale=.8]{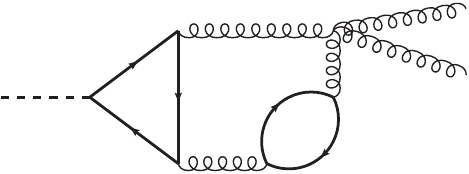}
    \end{subfigure}%
        \begin{subfigure}[b]{.35\textwidth}
        \centering
        \includegraphics[scale=.8]{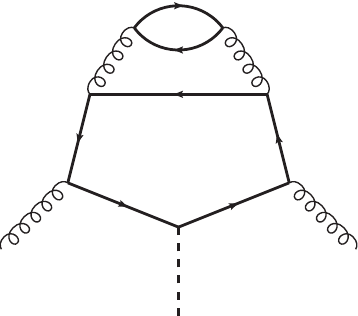}
    \end{subfigure}%
        \begin{subfigure}[b]{.35\textwidth}
        \centering
        \includegraphics[scale=.8]{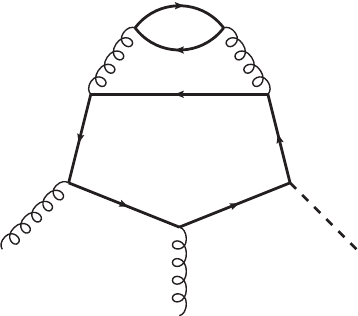}
    \end{subfigure}%
\caption{Complete set of Feynman diagrams with two fermion
  loops contributing to the Higgs-gluon form factor at three-loop
  order. The fermion loop connected to the Higgs-boson line
  corresponds to a massive quark. The quark of the second fermion loop
  may be either massive or massless.}
\label{fig:diags}
\end{figure}

In pure QCD, the evaluation of the form factor is complicated by the
fact that the process is loop induced. Nevertheless, exact two-loop
results for arbitrary quark masses have been available since
Refs.~\cite{Spira:1995rr, Harlander:2005rq, Anastasiou:2006hc,
Aglietti:2006tp}. Improvement over the current accuracy of cross
section predictions requires the knowledge of the form factor at
three-loop order. This is quite a challenging problem that has been
first attacked with the help of the large-mass expansion in the
top-quark mass \cite{Harlander:2009bw, Pak:2009bx}. A large-mass
expansion has even been derived at four-loop order
\cite{Davies:2019wmk}. Further progress at three-loops has been
recently achieved using Pad\'e approximants \cite{Davies:2019nhm}
exploiting partial knowledge of the form factor's behaviour around
threshold \cite{Grober:2017uho}. While a complete result for the form
factor at this order remains elusive, an exact result in terms of
harmonic polylogarithms has been obtained for contributions involving
a massless-quark loop \cite{Harlander:2019ioe}. The diagrams
contributing to the latter calculation are depicted in
Fig.~\ref{fig:diags}. The same diagrams also contribute with two
massive quark loops. In the present publication, we present an exact
result for the form factor in QCD with a single massive quark. In
particular, we compute the diagrams of Fig.~\ref{fig:diags} with both
quark loops with the same flavour, as well as the complete set of
diagrams with only one massive-quark loop. A result in QCD with
several massive quarks would still require a calculation of the
diagrams Fig.~\ref{fig:diags} with massive quarks of different
flavour.

Our results are certainly necessary to answer the question whether
Pad\'e approximants are indeed sufficient phenomenologically as
claimed in Ref.~\cite{Davies:2019nhm}. Independently, the knowledge of
exact quark mass dependence of the form factor opens the possibility
of including b-quark mass effects exactly.

The paper is organised as follows. In the next section, we introduce
our conventions and define finite remainders of the form factor after
infrared renormalisation. We use this opportunity to provide explicit
formulae for the scale dependence of the form factor as well. We
subsequently describe the methodology that has allowed us to obtain
not only a high precision numerical result but also high-order expansions
around the three physical singularities: infinite quark mass
(large-mass expansion), intermediate-quark production threshold
(threshold expansion) and vanishing quark
mass (high-energy expansion). Finally, we present our results and
compare them to previous work, in particular, the Pad\'e approximants of
Ref.~\cite{Davies:2019nhm}. This main text is closed with conclusions
and outlook. The three expansions are reproduced in separate
appendices. The last appendix presents the contents of an ancillary
file that contains our results in electronic form.


\section{Finite remainders}

Consider the amplitude for the fusion of two gluons of momenta
$p_{1,2}$, helicities $\lambda_{1,2}$ and adjoint-representation
colors $a_{1,2}$, followed by the production of one, possibly
off-shell, Higgs boson:
\begin{multline} \label{eq:amplitude}
-i \mathcal{M}\big[ g(p_1,\lambda_1,a_1) + g(p_2,\lambda_2,a_2) \to H
\big] \equiv \\[.2cm]
i \delta^{a_1a_2} \big[ (\epsilon_1 \cdot p_2) \, (\epsilon_2
\cdot p_1) - (\epsilon_1 \cdot \epsilon_2)  \, (p_2\cdot p_1) \big] \,
{1\over v} \frac{\alpha_s}{\pi} \, \mathcal{C} \; .
\end{multline}
Here, $v$ is the Higgs-doublet Vacuum Expectation Value. The coupling
of a single quark field, $Q$, of mass $M \neq 0$ to the Higgs-boson
field, $H$, is given by the tree-level Lagrangian term $-M \bar{Q} Q
H/v$. Finally, the gluon polarisation vectors are normalised as
follows:
\be
\epsilon_i \equiv \epsilon(\bm{p}_i,\lambda_i) \; , \qquad
\epsilon_i \cdot p_i = 0 \; , \qquad
\epsilon_i \cdot \epsilon^*_i = -1 \; , \qquad
i = 1,2 \; .
\ee
The {\bf Form Factor} $\mathcal{C}$ is expanded in the strong coupling
constant, $\alpha_s$, and the number of massless quark flavors, $n_l$:
\be \label{eq:expansion}
\begin{split}
\mathcal{C} &= \mathcal{C}^{(0)} + \frac{\alpha_s}\pi
\mathcal{C}^{(1)} + \left(\frac{\alpha_s}\pi\right)^2
\mathcal{C}^{(2)} + \order{\alpha_s^3} \; , \qquad
\mathcal{C}^{(n)} = \sum_{k=0}^n \mathcal{C}^{(n,k)} \, n_l^k \; .
\end{split}
\ee
The strong coupling is defined in the $\overline{\mathrm{MS}}$ scheme with
massive-quark decoupling. Its dependence on the renormalisation scale
$\mu$ is given by the $\beta$-function for $n_l$ massless quarks,
$\alpha_s \equiv \alpha_s^{(n_l)}(\mu)$. Contributions
$\mathcal{C}^{(n,n)} \neq 0$, $n > 0$ are only due to coupling
constant renormalisation. The massive-quark mass, $M$, is defined in
the on-shell scheme implying the same for the Yukawa coupling. The
dimensionless form-factor expansion coefficients depend on two
variables only:
\begin{align}
\mathcal{C}^{(n,k)} &\equiv \mathcal{C}^{(n,k)}\left( z , \, L_\mu
\right) \; , \\[.2cm]
z \equiv \frac{s}{4M^2}+i0^+ \; , \qquad
L_\mu &\equiv \ln\left( -\frac{\mu^2}{s + i0^+} \right) \; , \qquad
s \equiv (p_1 + p_2)^2 \; .
\end{align}
The leading contribution is:
\be
\mathcal{C}^{(0)} = \mathcal{C}^{(0,0)}= T_F {1\over z} \left\{ 1 -
  \left( 1 - {1\over z} \right) \left[ {1\over2} \ln\left(
      \frac{\sqrt{1-1/z}-1}{\sqrt{1-1/z}+1} \right) \right]^2 \right\}
\; .
\ee
In the limit $M \to \infty$:
\be
\mathcal{C}^{(0)}\big[z = 0\big] = {1\over3} \; .
\ee
Hence, the amplitude Eq.~\eqref{eq:amplitude} may be obtained at $M
\to \infty$ from the Higgs-Effective-Theory tree-level Lagrangian:
\be
\mathcal{L}^{(0)}_{\mathrm{HET}} = \frac{\alpha_s}{12\pi}
\frac{H}{v} \, G^{a}_{\mu\nu} G^{a \, \mu\nu} \; ,
\ee
where $G^a_{\mu\nu}$ is the standard QCD field-strength tensor,
$\mathcal{L}_{\mathrm{QCD}} = -1/4 \, G^{a}_{\mu\nu} G^{a \, \mu\nu} +
\mathcal{L}_{\mathrm{matter}}$.

Beyond leading order, the form factor is infrared divergent after
renormalisation. The results presented in this publication correspond
to Conventional Dimensional Regularisation with space-time dimension
$d = 4-2\epsilon$. The infrared divergences may be factorised yielding
the {\bf Finite Remainder}, $\mathcal{C}_I$, of the form factor:
\be \label{eq:finiteRemainderI}
\mathcal{C}_I \equiv I \, \mathcal{C} \; ,
\ee
where the two-loop $I$-operator of Catani \cite{Catani:1998bh} (see
Ref.~\cite{deFlorian:2012za} for the specific case of the Higgs-gluon
form factor) is given by:
\be \label{eq:Ioperator}
\begin{aligned}
I &= 1 - \frac{\alpha_s}{2\pi} I^{(1)} - \left( \frac{\alpha_s}{2\pi}
\right)^2 I^{(2)} \; , \\[.2cm]
I^{(1)} &\equiv I^{(1)}(\epsilon) =
-\left(-\frac{\mu^2}{s+i0^+}\right)^{\epsilon}
\frac{e^{\epsilon\gamma_E}}{\Gamma(1-\epsilon)}
\left[\frac{C_A}{\epsilon^2} + \frac{b_0}{2\epsilon}\right] \; ,
\\[.2cm]
I^{(2)} &= - \frac12 I^{(1)}(\epsilon) \left( I^{(1)}(\epsilon) +
  \frac{b_0}{\epsilon} \right) + \frac{e^{-\epsilon\gamma_E}
  \Gamma(1-2\epsilon)}{\Gamma(1-\epsilon)} \left(
  \frac{b_0}{2\epsilon} + K \right) I^{(1)}(2\epsilon) \\[.2cm]
&\quad+ \left(-\frac{\mu^2}{s+i0^+}\right)^{2\epsilon}
\frac{e^{\epsilon\gamma_E}}{\Gamma(1-\epsilon)} \frac{H_g}{2\epsilon}
\; ,
\end{aligned}
\ee
with the first two coefficients of the QCD $\beta$-function:
\be
b_0 = \frac{11}{3} C_A - \frac{4}{3} T_F n_l \; , \qquad
b_1 = \frac{34}{3} C_A^2 - \frac{20}{3} C_A T_F n_l - 4 C_F T_F n_l \; ,
\ee
and:
\be
\begin{aligned}
K &= \left( \frac{67}{18} - \frac{\pi^2}6 \right) C_A - \frac{10}{9} T_F
n_l \; , \\[.2cm]
H_g &= \left( \frac5{12} + \frac{11\pi^2}{144} + \frac{\zeta_3}2
\right) C_A^2 + \left( - \left( \frac{58}{27} + \frac{\pi^2}{36} \right) C_A
+ C_F + \frac{20}{27} T_F n_l \right) T_F n_l \; .
\end{aligned}
\ee
In general, $\mathcal{C}^{(n,n)}_I \neq 0$, $n > 0$. However:
\be \label{eq:nlDependence}
\mathcal{C}_I^{(1,1)}\big[ L_\mu = 0 \big] = 0 \; , \qquad
\mathcal{C}_I^{(2,2)}\big[ L_\mu = 0 \big] = \frac{\pi^2}{864}
\mathcal{C}^{(0)} \; .
\ee
Just as the form factor itself, the $I$-operator,
Eq.~\eqref{eq:Ioperator}, is independent of the scale $\mu$ (up to
two-loop order of course). In consequence:
\be
\dv{\ln \mathcal{C}_I}{\ln \mu} = \dv{\ln I}{\ln \mu} + \dv{\ln
  \mathcal{C}}{\ln \mu} = 0 \; .
\ee
The dependence of the finite remainder on the scale logarithm,
$L_\mu$, is thus given by the $\beta$-function only\footnote{Notice
that the $I$-operator of Ref.~\cite{Harlander:2019ioe} (see Eq.~(3.7b)
of that publication) is missing a scale-dependent factor in the
$H_g$-term (compare to Eq.~(4.38) of
Ref.~\cite{deFlorian:2012za}). With this difference, the $I$-operator
of Ref.~\cite{Harlander:2019ioe} is not scale invariant and
$\mathcal{C}_I^{(2)}$ contains an additional contribution to the
single scale-logarithm term, $H_g/4 \, \mathcal{C}^{(0)} L_\mu$.}:
\be \label{eq:scaleDependence}
\begin{aligned}
\mathcal{C}_I^{(1)} &= \mathcal{C}_I^{(1)}\big[ L_\mu = 0 \big] +
\frac{b_0}{4} \mathcal{C}^{(0)} \, L_\mu \; , \\[.2cm]
\mathcal{C}_I^{(2)} &= \mathcal{C}_I^{(2)}\big[ L_\mu = 0 \big] +
\frac{b_0}{2} \mathcal{C}_I^{(1)}\big[ L_\mu = 0 \big] \, L_\mu +
\frac{b_1 + b_0^2 L_\mu}{16} \, \mathcal{C}^{(0)} \, L_\mu \; .
\end{aligned}
\ee
A different finite remainder, $\mathcal{C}_Z$, is obtained if the
factorisation of infrared divergences is performed in the
$\overline{\mathrm{MS}}$ scheme \cite{Becher:2009cu}. Define:
\be \label{eq:finiteRemainderZ}
\mathcal{C}_Z \equiv Z^{-1} \, \mathcal{C} \; ,
\ee
with:
\be
\dv{\ln Z^{-1}}{\ln \mu} \equiv \Gamma \equiv -C_A
\gamma_{\mathrm{cusp}} L_\mu + 2 \gamma_g \; .
\ee
The solution at two-loops is:
\begin{align}
&\ln Z^{-1} = -\frac{\alpha_s}{4\pi} \left(
  \frac{\Gamma_0^\prime}{4\epsilon^2} + \frac{\Gamma_0}{2\epsilon}
\right) - \left( \frac{\alpha_s}{4\pi} \right)^2 \left( - \frac{3b_0
    \Gamma_0^\prime}{16\epsilon^3} + \frac{\Gamma_1^\prime -
    4b_0\Gamma_0}{16\epsilon^2} + \frac{\Gamma_1}{4\epsilon} \right)
\; , \\[.2cm]
&\Gamma^\prime \equiv \pdv{\Gamma}{\ln \mu} = -2C_A
\gamma_{\mathrm{cusp}} \; , \qquad
\Gamma \equiv \frac{\alpha_s}{4\pi} \, \Gamma_0 + \left(
  \frac{\alpha_s}{4\pi} \right)^2 \, \Gamma_1 \; ,
\end{align}
with the anomalous dimensions:
\be
\begin{aligned}
&\gamma_{\mathrm{cusp}} = \frac{\alpha_s}{\pi} + \left(
  \frac{\alpha_s}{\pi} \right)^2 \frac{K}{2} \; , \\[.2cm]
&\gamma_g = - \frac{\alpha_s}{4\pi} b_0 + \left( \frac{\alpha_s}{4\pi}
\right)^2 \left[ \left( - \frac{692}{27} + \frac{11 \pi^2}{18} + 2
    \zeta_3 \right) C_A^2 + \left( \left( \frac{256}{27} - \frac{2
        \pi^2}{9} \right) C_A + 4 C_F \right) T_F n_l \right] \; .
\end{aligned}
\ee
Since the dependence on the highest-power of $n_l$ in
Eq.~\eqref{eq:expansion} is only due to the pure poles in the
minimal ultraviolet renormalisation constant $Z_{\alpha_s}$, it must
be cancelled by the, equally minimal, constant $Z$. Thus:
\be
\mathcal{C}^{(n,n)}_Z = 0 \; .
\ee
The scale dependence of $\mathcal{C}_Z$, on the other hand, is
non-trivial:
\be
\dv{\ln C_Z}{\ln \mu} = \dv{\ln Z^{-1}}{\ln \mu} + \dv{\ln
  \mathcal{C}}{\ln \mu} = \Gamma \; .
\ee
The conversion between the two infrared schemes is achieved with the
help of:
\be
\mathcal{C}_Z = \left( IZ \right)^{-1} \mathcal{C}_I \; .
\ee
Explicitly:
\be \label{eq:conversion}
\begin{split}
\left( IZ \right)^{-1} &= 1
+ \frac{\alpha_s}{\pi} \left\{
\frac{\pi^2}{24} C_A
+ \left[ -\frac{11}{12} C_A + \frac{1}{3} T_F n_l \right] L_\mu
- \frac{1}{4} C_A L_\mu^2
\right\} \\[.2cm]
&\quad+ \left( \frac{\alpha_s}{\pi} \right)^2 \left\{
- \left( \frac{\pi^2}{64} + \frac{11\zeta_3}{96} \right) C_A^2
+ \left( \left( \frac{17\pi^2}{864} + \frac{\zeta_3}{24} \right) C_A -
  \frac{\pi^2}{216} T_F n_l \right) T_F n_l
\right. \\[.2cm]
&\qquad \qquad \qquad \!
+ \left[ \left( -\frac{173}{108} + \frac{11\pi^2}{288} +
    \frac{\zeta_3}{8} \right) C_A^2 + \left( \left( \frac{16}{27} -
      \frac{\pi^2}{72} \right) C_A + \frac{1}{4} C_F \right) T_F n_l
\right] L_\mu \\[.2cm]
&\qquad \qquad \qquad \!
+ \left[ \left( -\frac{67}{144} + \frac{\pi^2}{96} \right) C_A^2 +
  \frac{5}{36} C_A T_F n_l \right] L_\mu^2
+ \left[ \frac{11}{72} C_A^2 - \frac{1}{18} C_A T_F n_l \right]
L_\mu^3 \\[.2cm]
&\qquad \qquad \qquad \! \left.
+ \frac{1}{32} C_A^2 L_\mu^4 \right\} \; .
\end{split}
\ee
For instance, this result allows to obtain
Eqs.~\eqref{eq:nlDependence} and the scale dependence of
$\mathcal{C}_Z$ after using Eqs.~\eqref{eq:scaleDependence}.

Finally, let us note that our results can be used to obtain the
three-loop form factor before factorisation of the infrared divergences
with the help of the two-loop result provided to
$\order{\epsilon^2}$ in Ref.~\cite{Anastasiou:2020qzk}.


\section{Technicalities} \label{sec:technicalities}

The three-loop diagrams corresponding to the amplitude
Eq.~\eqref{eq:amplitude} have been reduced to a set of (master)
integrals, $M_i(z,\epsilon)$, via Integration-By-Parts
identities \cite{Chetyrkin:1981qh} with the help of a \textsc{C++}
implementation \cite{DiaGenIdSolver} of the Laporta algorithm
\cite{Laporta:2001dd}. The same reduction has also been exploited to
construct a system of first-order homogeneous linear differential
equations \cite{Kotikov:1990kg, Remiddi:1997ny}:
\be \label{eq:masterDEs}
\dv{M_i (z,\epsilon)}{z} \equiv \sum_j
A_{ij}(z,\epsilon) \, M_j(z,\epsilon) \; ,
\ee
where the coefficients $A_{ij}(z,\epsilon)$ are rational
functions in $z$ and $\epsilon$. Truncated $\epsilon$-expansions have
been subsequently substituted to represent the master integrals. A
large-mass expansion (see below) of each $M_i$ has been used to
determine the lowest power of $\epsilon$, $\underline{n}_i$,
with non-vanishing coefficient, while the amplitude and the
differential equations have been used to determine the highest
power of $\epsilon$, $\overline{n}_i$, necessary to obtain the
amplitude at $\order{\epsilon^0}$. Let the coefficients of the
truncated $\epsilon$-expansions be denoted with $I_k(z)$:
\be
M_i (z,\epsilon) \equiv \sum_{l =
  0}^{\overline{n}_i-\underline{n}_i} \epsilon^{\underline{n}_i + l}
\, I_{\underline{k}_i + l}(z) \; ,
\ee
where $\underline{k}_i$ have been chosen to avoid overlap of the
$k$-indices of the expansion coefficients $I_k$ of different master
integrals. The coefficients $I_k$ satisfy a system of first-order
homogeneous linear differential equations derived from
Eqs.~\eqref{eq:masterDEs}:
\be \label{eq:coeffDEs}
\dv{I_k (z)}{z} \equiv \sum_l B_{kl}(z) \,
I_l(z) \; ,
\ee
where the coefficients $B_{kl}(z)$ are rational functions in
$z$. Instead of seeking an analytic solution of
Eqs.~\eqref{eq:coeffDEs}, we have solved the system numerically as
proposed originally in Ref.~\cite{Caffo:1998du} and first applied to a
physical problem in Ref.~\cite{Boughezal:2007ny}. To this end,
we have used the \textsc{Boost} \cite{Boost} library \verb|odeint|. In
particular, we have chosen the Bulirsch-Stoer algorithm,
\verb|bulirsch_stoer_dense_out|. In order to keep the numerical
precision of the results under control, we have used the
\textsc{Boost} library \verb|multiprecision| with a
\verb|gmp|/\verb|mpc| backend. The floating point containers were
requested to represent 100 decimal digits. A local error of $10^{-40}$
has been requested from the differential equation solution.

\begin{figure}[t]
\includegraphics[width=\textwidth]{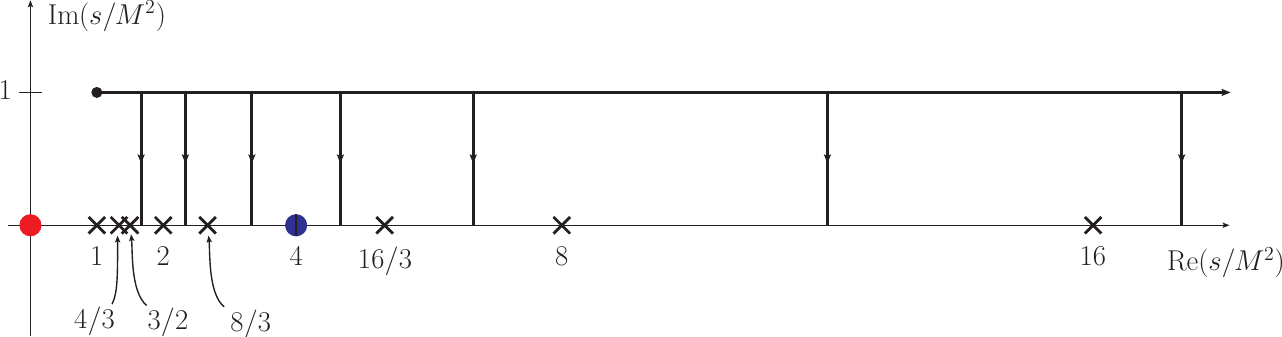}
\caption{Contours for the numerical solution of the differential
  equations for the master integrals. The points on the abscissa
  correspond to singularities of the differential equations. Every time
  a contour reaches the real axis, the interval between singularities is
  explored in both directions.}
\label{fig:contour}
\end{figure}

The numerical solution of Eqs.~\eqref{eq:coeffDEs} requires a boundary
value for each $I_k$. In order to obtain these, we have used
a high-order large-mass expansion, see e.g.\ \cite{Smirnov:2002pj},
around $z = 0$. The expansion must have unit radius of
convergence\footnote{Strictly speaking, this is a power-log expansion
with singularity at $z = 0$. The convergence considerations apply to
the coefficients of the logarithms, $\ln^m z$, which are analytic in
$z$.}  in $z$, since the nearest singularity of the master integrals
is at $z = 1$. The expansion has been obtained using diagrammatic
methods for the first few coefficients. It has been subsequently
extended with the help of the differential equations. As boundary
point, we have chosen $z = 1/4(1+i)$, well within the radius of
convergence. Because of the presence of singularities in the
coefficients $B_{kl}$, we have used evolution contours shown
in Fig.~\ref{fig:contour}. An additional solution has also been
obtained starting from $z = 1/4(0.7+0.7i)$ in order to control the
error of the final result.

Having high-precision values of the master integrals allows to obtain
expansions around arbitrary points, even around singularities. In the
course of the present work, we have obtained threshold and
high-energy expansions. They are necessary to evaluate the three-loop
coefficient of the form factor in the vicinity of $z = 1$ and $1/z =
0$ respectively. In general, expansions of $I_k$ are of
power-log type, since an expansion in $\epsilon$ of the master
integrals has already been performed:
\be
I_k\big( z(y) \big)  \equiv \sum_{l = \underline{l}_k}^\infty \sum_{m =
  \underline{m}_k}^{\overline{m}_k} c_{klm} \, y^l \ln^m y \; ,
\ee
where $\underline{l}_k, \underline{m}_k, \overline{m}_k \in
\mathbb{Z}$, and $y = \sqrt{1-z}$ for the threshold expansion, while
$y = 1/z$ for the high-energy expansion. In practice, the expansions
are truncated at an affordable order considering the available
computing ressources. For each $I_k$, only one $c_k \equiv c_{klm}$
for some $l$ and $m$, is necessary to make the solution of
Eqs.~\eqref{eq:coeffDEs} unique. Since Eqs.~\eqref{eq:coeffDEs} are
linear, there is:
\be \label{eq:matching}
I_k\big( z(y) \big) \equiv \sum_l F_{kl}(y) \, c_l
\qquad \Longrightarrow \qquad
c_k = \sum_l \big( F^{-1} \big)_{kl}(y) \, I_l\big( z(y) \big) \; .
\ee
In order to obtain $c_{klm}$ and thus also $F_{kl}(y)$, we have used
an efficient \textsc{C++} software that was originally developed for
Ref.~\cite{Czakon:2015exa}. Upon choosing a suitable $y$ point where
the threshold or the high-energy expansion has excellent convergence,
we were able to obtain $c_k$ with high precision.


\section{Results}

Since the scale logarithms of the three-loop coefficient of the finite
remainder are entirely determined from the analytically known lower
order results, see Eqs.~\eqref{eq:scaleDependence}, we only present
our findings at $L_\mu = 0$.

\begin{figure}[t]
\center
\includegraphics[width=0.49\textwidth]{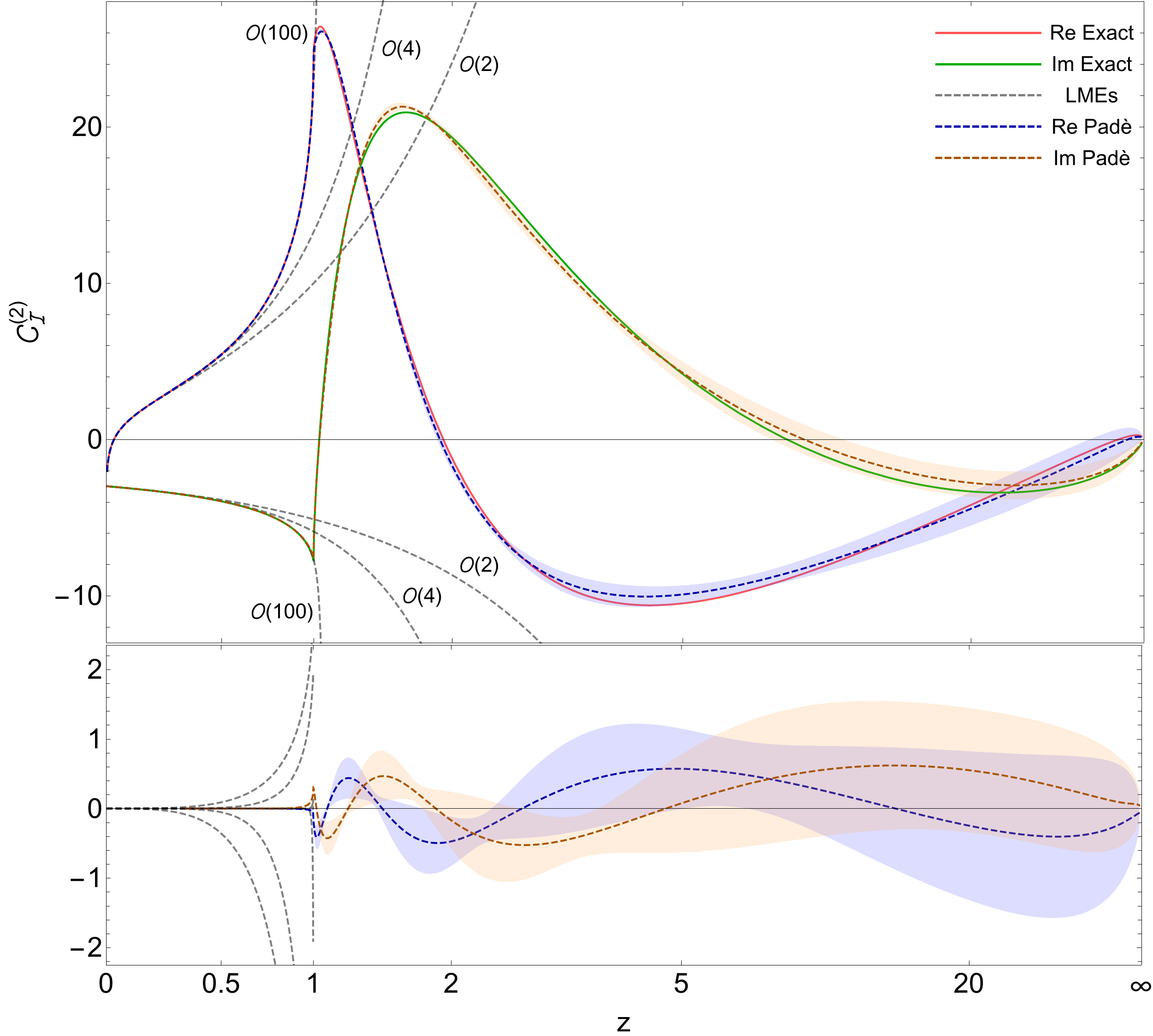}
\includegraphics[width=0.49\textwidth]{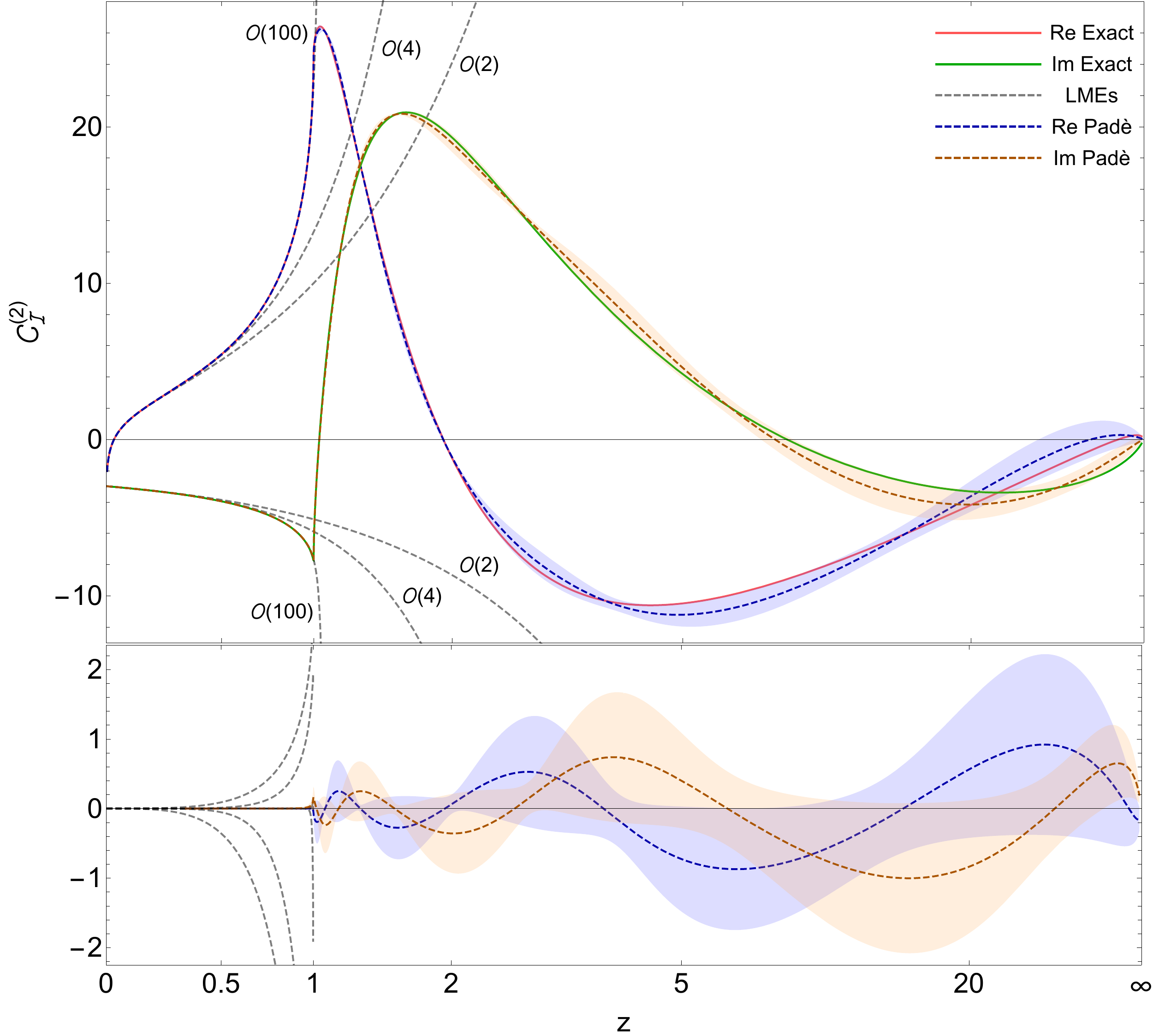}
\caption{Comparison of the three-loop coefficient of the finite
  remainder, Eq.~\eqref{eq:finiteRemainderI}, at $n_l = 5$, $L_\mu = 0$
  (five massless quarks, renormalisation scale $\mu^2 = -s$), with the
  default Pad\'e approximation, $[6,1]$, constructed in
  Ref.~\cite{Davies:2019nhm} (left panel) and improved to $[7,1]$ in
  Ref.~\cite{Davies:2019roy} (right panel), as function of $z = s/4M^2$
  with $\sqrt{s}$ the center-of-mass energy of the Higgs boson and $M$
  the mass of the single massive quark. The bands correspond to the
  uncertainty of the Pad\'e approximations as estimated in
  Refs.~\cite{Davies:2019nhm}~and~\cite{Davies:2019roy}. The lower plot
  shows the absolute difference between the approximation and the exact
  result. Also shown is the large-mass expansion (LME) of the three-loop
  coefficient of the finite remainder truncated at $\order{z^2},
  \order{z^4}$ and $\order{z^{100}}$.}
\label{fig:comparisonNL5}
\end{figure}

\begin{figure}[t]
\center
\includegraphics[width=0.49\textwidth]{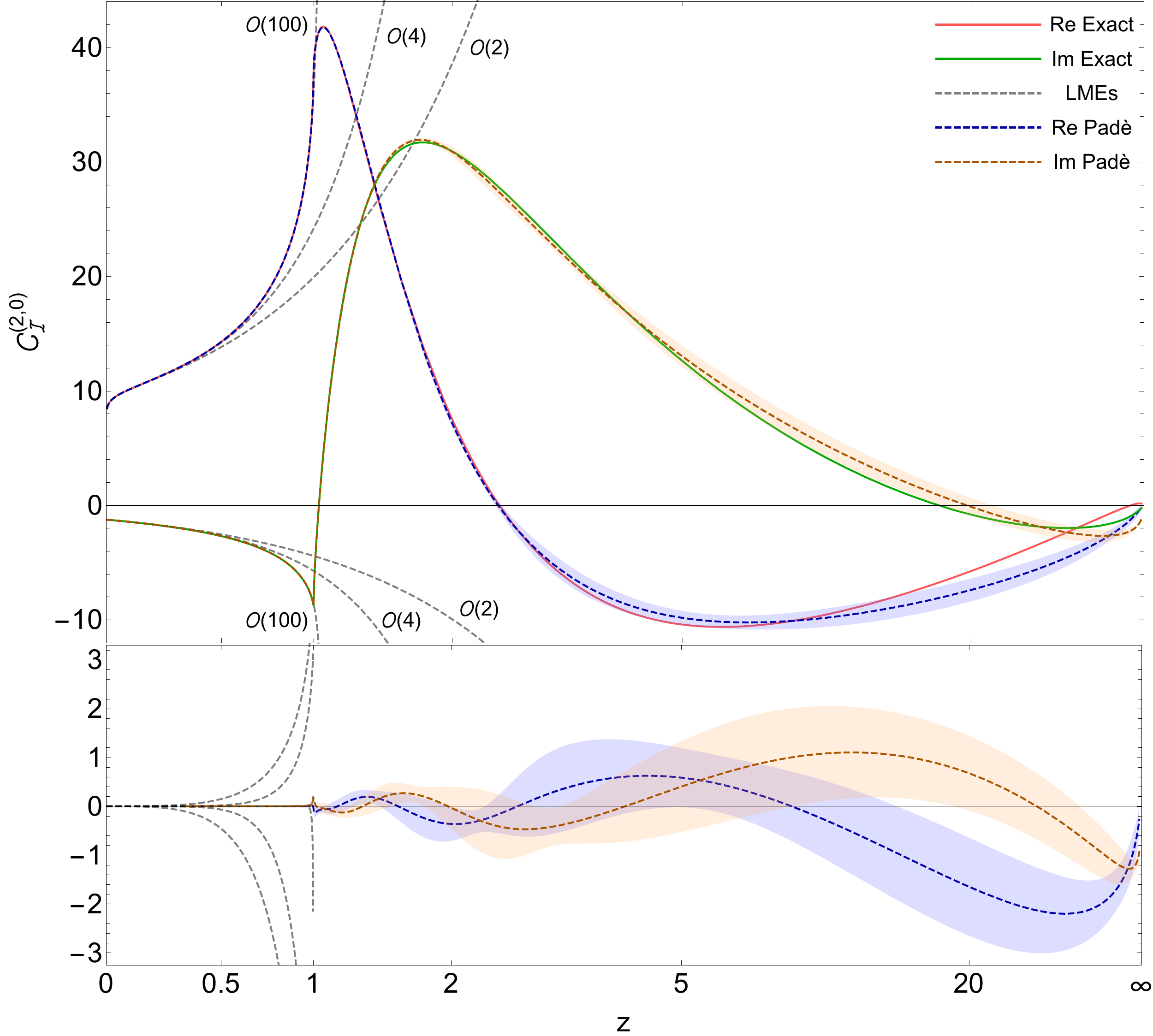}
\includegraphics[width=0.49\textwidth]{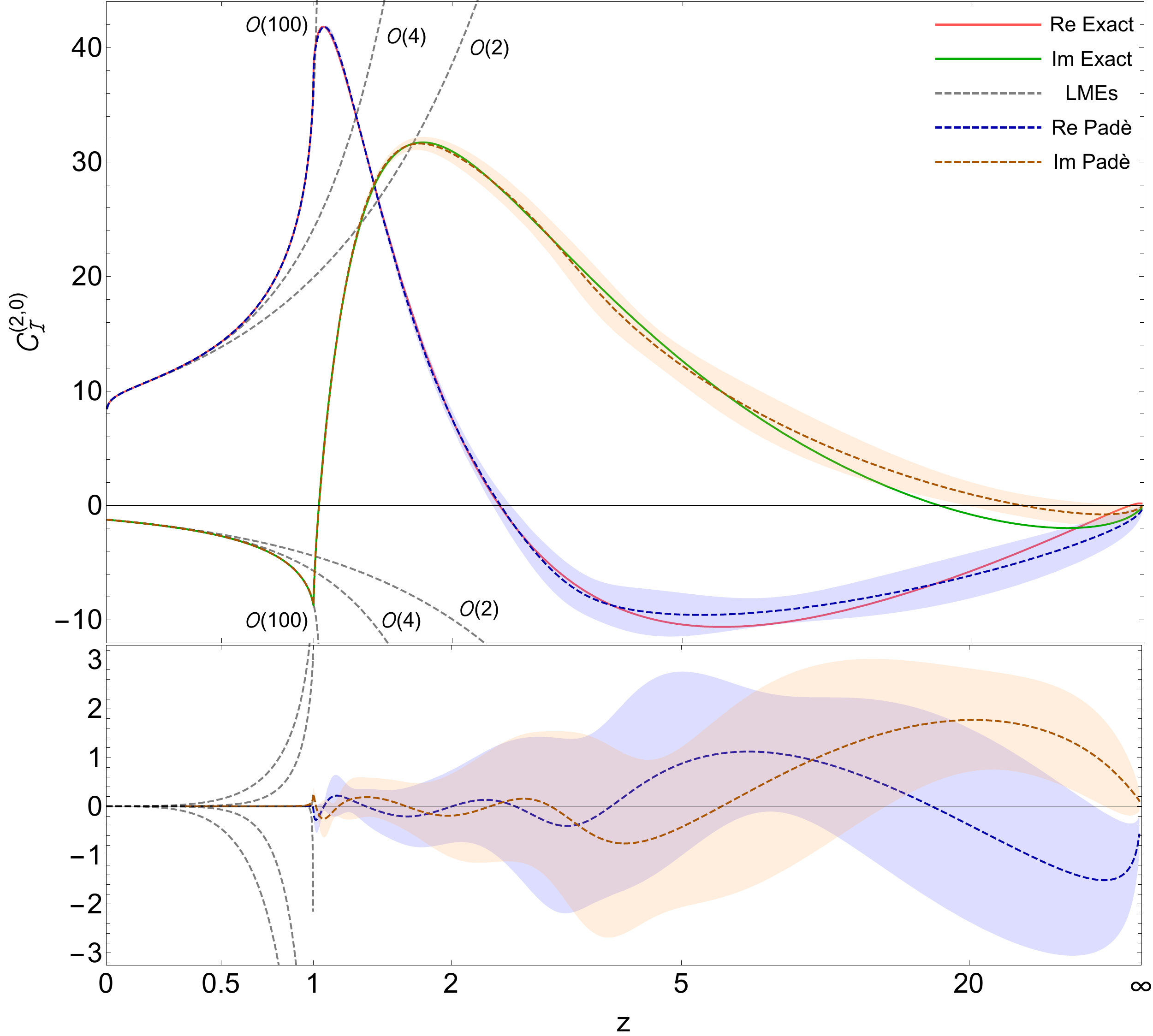}
\caption{Same as Fig.~\ref{fig:comparisonNL5} but with $n_l = 0$.}
\label{fig:comparisonNL0}
\end{figure}

We first note that our result for $\mathcal{C}_I^{(2,1)}$ agrees
perfectly with Ref.~\cite{Harlander:2019ioe}. Remains to compare
with the Pad\'e approximants of Ref.~\cite{Davies:2019nhm} for
$\mathcal{C}^{(2)}$. A comparison for the case of five massless quarks
is presented in Fig.~\ref{fig:comparisonNL5}. We observe that the
uncertainty estimates of the approximants are reliable over most of
the range of $z$. Slightly larger deviations are observed for the $n_l
= 0$ case as demonstrated in Fig.~\ref{fig:comparisonNL0}. An
improvement of the Pad\'e approximants has recently appeared in the
proceedings \cite{Davies:2019roy}. The respective plots are also shown
in Figs.~\ref{fig:comparisonNL5}~and~\ref{fig:comparisonNL0}. Clearly,
the agreement with the exact result is worse for $n_l = 5$ and better
for $n_l = 0$.

\begin{figure}[b]
\center
\includegraphics[width=0.49\textwidth]{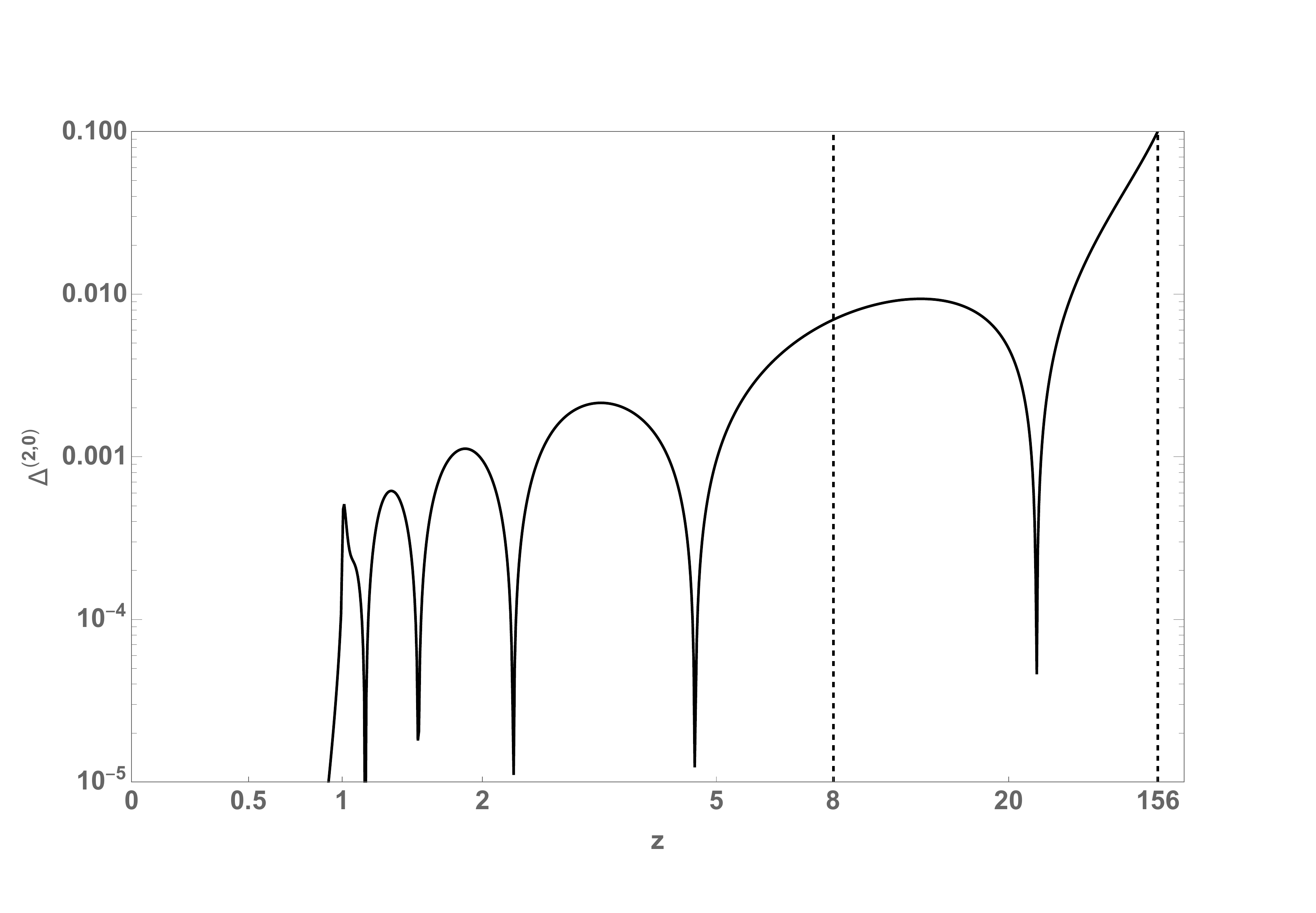}
\includegraphics[width=0.49\textwidth]{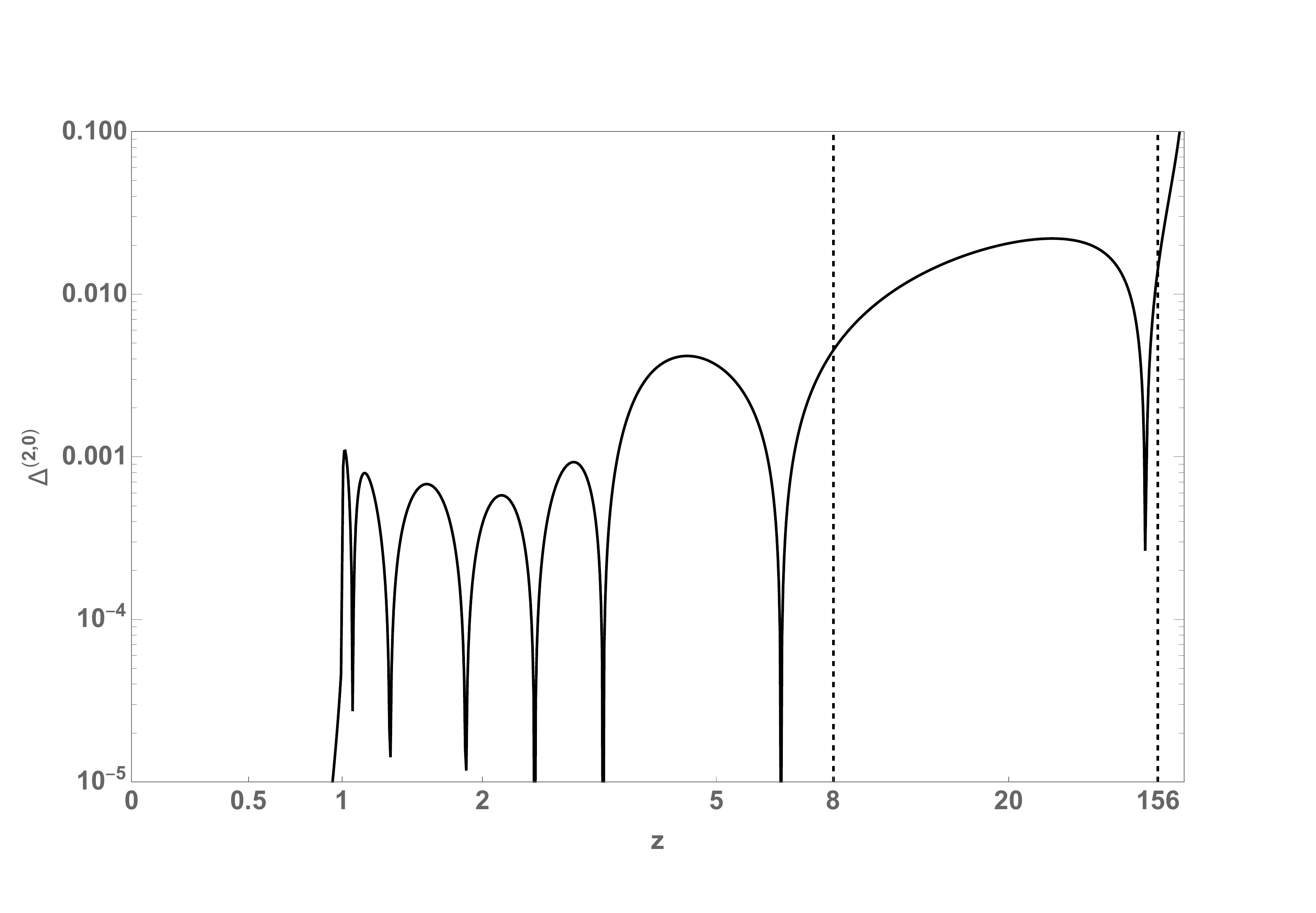}
\caption{Relative difference, Eq.~\eqref{eq:Delta}, between the Pad\'e
  approximation of the three-loop coefficient of the finite remainder
  $\mathcal{C}_I^{(2)}$ from Refs.~\cite{Davies:2019nhm} (left panel)
  and \cite{Davies:2019roy} (right panel) and the exact result at $n_l =
  0$, $L_\mu = 0$. $z \approx 8$ corresponds to a $\sqrt{s}$ = 1~TeV
  Higgs boson produced through a top-quark loop, whereas $z \approx 156$
  corresponds to an on-shell Higgs boson produced through a b-quark
  loop.}
\label{fig:Delta}
\end{figure}

In order to understand the phenomenological relevance of the
difference between the exact result and its Pad\'e approximation for
$n_l = 0$, we consider the quantity:
\be \label{eq:Delta}
\Delta^{(2,0)} \equiv \left| \left( \frac{\alpha_s}{\pi} \right)^2
\frac{2\Re\left[ \left(
      \eval{\mathcal{C}_I^{(2,0)}}_{[6,1]-\mathrm{Pad\acute{e}}} -
      \mathcal{C}_I^{(2,0)} \right) \, \mathcal{C}^{(0)}
  \right]}{\left| \mathcal{C}^{(0)} \right|^2} \right| \; ,
\ee
as a proxy for the error induced on the partonic cross section. We
acknowledge the limitations of $\Delta^{(2,0)}$ in this respect due to
the size of the real-radiation corrections to the cross section at
higher orders. We expect that the actual effect is about 1/2 of
$\Delta^{(2,0)}$, at least for a top-quark loop. For simplicity, we
fix the value of the strong coupling at $\alpha_s =
0.1$. $\Delta^{(2,0)}$ is plotted in Fig.~\ref{fig:Delta}. Assuming an
off-shell Higgs-boson with a partonic center-of-mass energy,
$\sqrt{s}$, of up to 1~TeV produced through a top-quark loop, there is
$\Delta^{(2,0)} < 1 \%$. Hence, the Pad\'e approximant provides an
excellent approximation for top-quark loops. On the other hand, in the
case of the production of an on-shell Higgs boson through a b-quark
loop, $\Delta^{(2,0)} \approx 10 \%$.  Furthermore, the difference
grows rapidly with the Higgs-boson off-shellness, $\sqrt{s}$. Hence,
the approximation is rather poor for b-quark loops. In the same
figure, we also show $\Delta^{(2,0)}$ using the improved Pad\'e
approximant of Ref.~\cite{Davies:2019roy}. We note that the
approximation is now better for b-quarks. Nevertheless,
$\Delta^{(2,0)} > 10\%$  for an off-shell Higgs boson of 400~GeV.

\begin{figure}[t]
\center
\includegraphics[width=\textwidth]{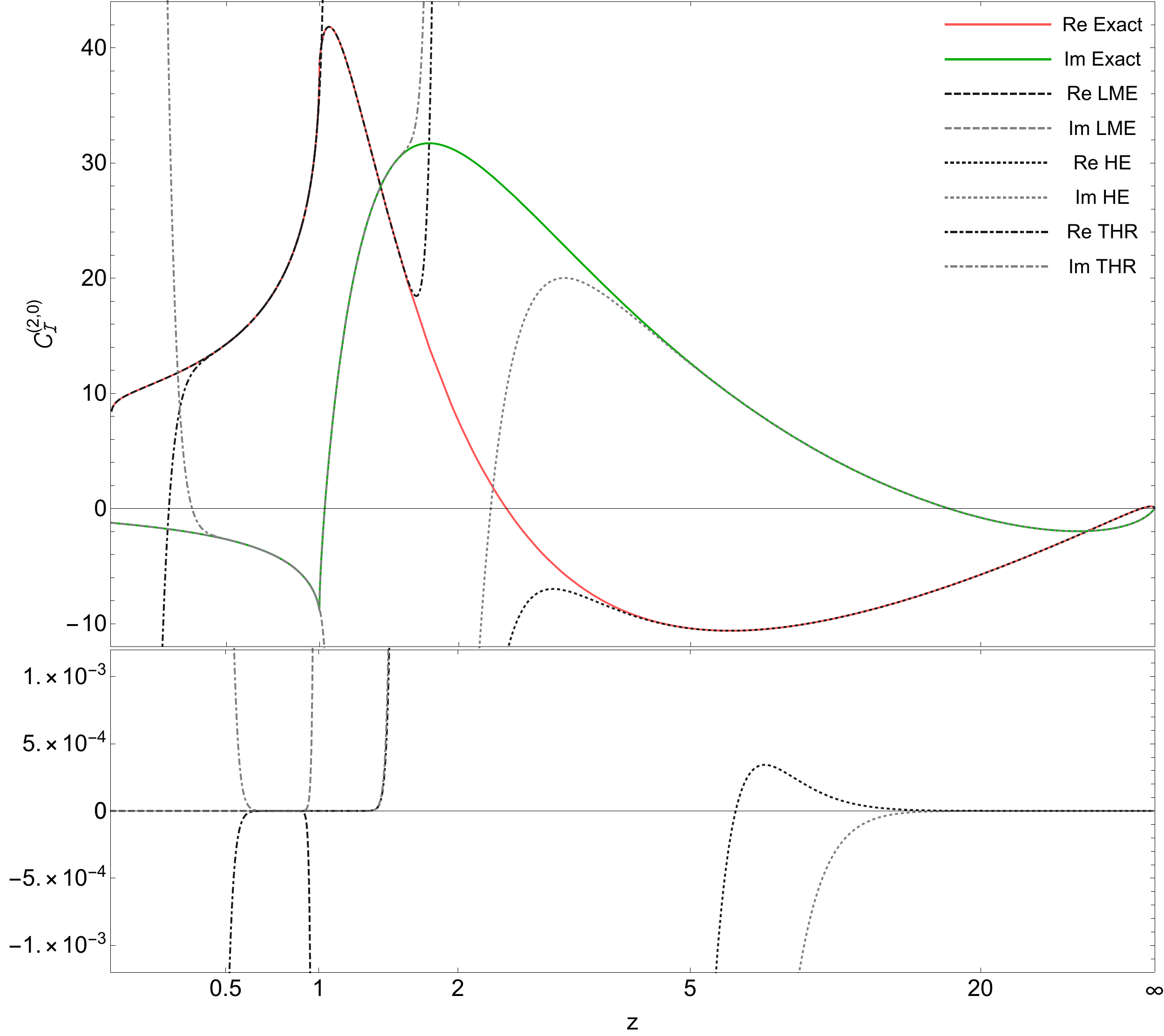}
\caption{Comparison of the large-mass expansion (LME) truncated at
  $\order{z^{100}}$, threshold expansion (THR) truncated at
  $\order{(1-z)^{20}}$ and high-energy expansion (HE) truncated at
  $\order{1/z^{8}}$ with the exact result for the three-loop
  coefficient of the finite remainder $\mathcal{C}_I^{(2)}$ at $n_l =
  0$, $L_\mu = 0$. The lower panel shows the absolute difference
  between the expansions and the exact result.}
\label{fig:best}
\end{figure}

Our exact result is a sample of $\mathcal{C}_I^{(2)}$ values at nearly
200.000 $z$ points. We have also determined the large-mass, threshold
and high-energy expansions of $\mathcal{C}_I^{(2)}$ (see
Section~\ref{sec:technicalities}). These three expansions cover most of
the range of $z$ values within their convergence radii. In the
Supplemental Material \cite{ggHCzakonNiggetiedt} (see
Appendix~\ref{sec:supplemental}) to the present publication, we
provide the large-mass expansion up to $\order{z^{100}}$ with exact
coefficients, the threshold expansion up to $\order{(1-z)^{20}}$ with
numerical coefficients and the high-energy expansion up to
$\order{1/z^{8}}$ with numerical coefficients. The order at which the
high-energy expansion has been truncated has been determined by
the requirement that the numerical expansion coefficients have at
least ten correct digits as determined in a conservative comparison of
results obtained with two different starting points for the numerical
solution of Eqs.~\eqref{eq:coeffDEs} and $y$ values
Eq.~\eqref{eq:matching}. The agreement of the truncated expansions
with the exact result is demonstrated in Fig.~\ref{fig:best}.

The domain of physical $z$ values may be compactified with the
following mapping:
\be
z(\rho) \equiv \frac{4\rho}{1 - \rho} \; , \qquad
\rho(z) = \frac{z}{4+z} \; , \qquad
\rho \in (0,1) \; .
\ee
The exact result for $\mathcal{C}_I^{(2,0)}$ is approximated to better
than $10^{-5}$ relative to $|\mathcal{C}^{(2,0)}|$ as follows:
\begin{center}
\begin{tabular}{ll}
$0 < \rho < 1/6$ & - large-mass expansion,
  Appendix~\ref{sec:LME} and Fig.~\ref{fig:comparisonNL0}; \\[.2cm]
$1/6 \leq \rho < 1/4$ & - threshold expansion,
  Appendix~\ref{sec:THR} and Fig.~\ref{fig:THR}; \\[.2cm]
$1/4 \leq \rho < 3/4$ & - interpolation of a sample of numerical values,
  Tabs.~\ref{tab:num1}~and~\ref{tab:num2}; \\[.2cm]
$3/4 \leq \rho < 1$ & - high-energy expansion,
  Appendix~\ref{sec:HE} and Fig.~\ref{fig:HE}.
\end{tabular}
\end{center}
\begin{table}[t]
\center
\begin{tabular}{|c|c|c|c|}
\hline
$\rho$ & $\mathcal{C}^{(2,0)}_I$ & $\rho$ & $\mathcal{C}^{(2,0)}_I$ \\
\hline
$1/4$
 & $30.88057646
+25.98752971 \, i
$ & $3/8$
 & $0.5489407632
+28.08768382 \, i
$ \\
$51/200$
 & $29.16117325
+27.19326399 \, i
$ & $19/50$
 & $-0.1268390632
+27.6738637 \, i
$ \\
$13/50$
 & $27.46093382
+28.21076656 \, i
$ & $77/200$
 & $-0.7713324763
+27.25087704 \, i
$ \\
$53/200$
 & $25.78986495
+29.06161664 \, i
$ & $39/100$
 & $-1.385714578
+26.82008886 \, i
$ \\
$27/100$
 & $24.15526667
+29.76456733 \, i
$ & $79/200$
 & $-1.971122667
+26.38273798 \, i
$ \\
$11/40$
 & $22.56238753
+30.33601069 \, i
$ & $2/5$
 & $-2.528655721
+25.93994889 \, i
$ \\
$7/25$
 & $21.01490693
+30.79034303 \, i
$ & $81/200$
 & $-3.05937427
+25.49274254 \, i
$ \\
$57/200$
 & $19.51529601
+31.14025791 \, i
$ & $41/100$
 & $-3.56430059
+25.04204591 \, i
$ \\
$29/100$
 & $18.06509163
+31.39698515 \, i
$ & $83/200$
 & $-4.044419136
+24.58870072 \, i
$ \\
$59/200$
 & $16.6651066
+31.5704884 \, i
$ & $21/50$
 & $-4.500677174
+24.13347121 \, i
$ \\
$3/10$
 & $15.31559266
+31.66963034 \, i
$ & $17/40$
 & $-4.933985559
+23.67705121 \, i
$ \\
$61/200$
 & $14.01636758
+31.70231201 \, i
$ & $43/100$
 & $-5.345219629
+23.22007047 \, i
$ \\
$31/100$
 & $12.76691514
+31.67559125 \, i
$ & $87/200$
 & $-5.735220182
+22.76310042 \, i
$ \\
$63/200$
 & $11.566464
+31.59578403 \, i
$ & $11/25$
 & $-6.104794506
+22.30665937 \, i
$ \\
$8/25$
 & $10.4140502
+31.46855168 \, i
$ & $89/200$
 & $-6.454717455
+21.85121724 \, i
$ \\
$13/40$
 & $9.308566879
+31.29897624 \, i
$ & $9/20$
 & $-6.785732545
+21.39719977 \, i
$ \\
$33/100$
 & $8.248803784
+31.0916259 \, i
$ & $91/200$
 & $-7.098553057
+20.94499247 \, i
$ \\
$67/200$
 & $7.233478837
+30.85061204 \, i
$ & $23/50$
 & $-7.393863147
+20.49494408 \, i
$ \\
$17/50$
 & $6.261263221
+30.57963911 \, i
$ & $93/200$
 & $-7.672318937
+20.04736981 \, i
$ \\
$69/200$
 & $5.330801353
+30.28204836 \, i
$ & $47/100$
 & $-7.934549597
+19.60255421 \, i
$ \\
$7/20$
 & $4.44072674
+29.96085629 \, i
$ & $19/40$
 & $-8.181158403
+19.16075384 \, i
$ \\
$71/200$
 & $3.589674492
+29.61878862 \, i
$ & $12/25$
 & $-8.412723764
+18.72219971 \, i
$ \\
$9/25$
 & $2.776291163
+29.2583102 \, i
$ & $97/200$
 & $-8.629800232
+18.28709941 \, i
$ \\
$73/200$
 & $1.999242412
+28.88165164 \, i
$ & $49/100$
 & $-8.832919461
+17.85563919 \, i
$ \\
$37/100$
 & $1.257218899
+28.49083281 \, i
$ & $99/200$
 & $-9.022591138
+17.42798575 \, i
$ \\

\hline
\end{tabular}
\caption{Numerical values of the three-loop coefficient of the finite
  remainder $\mathcal{C}_I^{(2)}$ at $n_l = 0$, $L_\mu = 0$, for $1/4
  \leq \rho \equiv z/(4+z) < 1/2$.}
\label{tab:num1}
\end{table}
\begin{table}[t]
\center
\begin{tabular}{|c|c|c|c|}
\hline
$\rho$ & $\mathcal{C}^{(2,0)}_I$ & $\rho$ & $\mathcal{C}^{(2,0)}_I$ \\
\hline
$1/2$
 & $-9.199303854
+17.00428794 \, i
$ & $5/8$
 & $-10.49655344
+7.904442944 \, i
$ \\
$101/200$
 & $-9.363525841
+16.58467833 \, i
$ & $63/100$
 & $-10.45672407
+7.602825895 \, i
$ \\
$51/100$
 & $-9.515705879
+16.16927502 \, i
$ & $127/200$
 & $-10.41175312
+7.305965242 \, i
$ \\
$103/200$
 & $-9.656274528
+15.75818283 \, i
$ & $16/25$
 & $-10.36179862
+7.013842682 \, i
$ \\
$13/25$
 & $-9.785644947
+15.35149431 \, i
$ & $129/200$
 & $-10.30701259
+6.726439296 \, i
$ \\
$21/40$
 & $-9.904213631
+14.94929071 \, i
$ & $13/20$
 & $-10.24754124
+6.443735699 \, i
$ \\
$53/100$
 & $-10.01236111
+14.55164295 \, i
$ & $131/200$
 & $-10.1835252
+6.165712192 \, i
$ \\
$107/200$
 & $-10.11045262
+14.15861246 \, i
$ & $33/50$
 & $-10.11509978
+5.892348901 \, i
$ \\
$27/50$
 & $-10.19883876
+13.77025201 \, i
$ & $133/200$
 & $-10.04239511
+5.623625905 \, i
$ \\
$109/200$
 & $-10.2778561
+13.38660649 \, i
$ & $67/100$
 & $-9.965536402
+5.359523369 \, i
$ \\
$11/20$
 & $-10.34782779
+13.00771356 \, i
$ & $27/40$
 & $-9.88464409
+5.100021655 \, i
$ \\
$111/200$
 & $-10.40906411
+12.63360429 \, i
$ & $17/25$
 & $-9.79983404
+4.845101447 \, i
$ \\
$14/25$
 & $-10.46186303
+12.2643038 \, i
$ & $137/200$
 & $-9.711217707
+4.594743853 \, i
$ \\
$113/200$
 & $-10.5065107
+11.89983176 \, i
$ & $69/100$
 & $-9.618902308
+4.34893052 \, i
$ \\
$57/100$
 & $-10.54328201
+11.54020288 \, i
$ & $139/200$
 & $-9.522990977
+4.107643733 \, i
$ \\
$23/40$
 & $-10.57244099
+11.18542744 \, i
$ & $7/10$
 & $-9.423582916
+3.870866519 \, i
$ \\
$29/50$
 & $-10.59424131
+10.83551164 \, i
$ & $141/200$
 & $-9.320773537
+3.638582749 \, i
$ \\
$117/200$
 & $-10.60892672
+10.49045807 \, i
$ & $71/100$
 & $-9.214654604
+3.410777238 \, i
$ \\
$59/100$
 & $-10.61673142
+10.15026603 \, i
$ & $143/200$
 & $-9.105314363
+3.187435844 \, i
$ \\
$119/200$
 & $-10.61788051
+9.814931857 \, i
$ & $18/25$
 & $-8.992837666
+2.968545567 \, i
$ \\
$3/5$
 & $-10.6125903
+9.484449303 \, i
$ & $29/40$
 & $-8.877306096
+2.754094652 \, i
$ \\
$121/200$
 & $-10.60106877
+9.158809768 \, i
$ & $73/100$
 & $-8.758798082
+2.54407269 \, i
$ \\
$61/100$
 & $-10.58351579
+8.838002595 \, i
$ & $147/200$
 & $-8.637389011
+2.338470728 \, i
$ \\
$123/200$
 & $-10.56012358
+8.52201532 \, i
$ & $37/50$
 & $-8.513151331
+2.137281371 \, i
$ \\
$31/50$
 & $-10.5310769
+8.210833902 \, i
$ & $149/200$
 & $-8.386154663
+1.940498901 \, i
$ \\
$5/8$
 & $-10.49655344
+7.904442944 \, i
$ & $3/4$
 & $-8.256465888
+1.748119392 \, i
$ \\

\hline
\end{tabular}
\caption{Numerical values of the three-loop coefficient of the finite
  remainder $\mathcal{C}_I^{(2)}$ at $n_l = 0$, $L_\mu = 0$, for $1/2
  \leq \rho \equiv z/(4+z) \leq 3/4$.}
\label{tab:num2}
\end{table}
%


\section{Conclusions and outlook}

With the results presented in this work, the Higgs-gluon form factor
is known exactly at three loops in QCD with a single massive
quark. This is sufficient for applications to Higgs-boson
hadroproduction in the five-flavour scheme, where the massive quark is
the top. In this case, we have confirmed that an approach based on
Pad\'e approximants \cite{Davies:2019nhm} is sufficient to obtain
sub-percent precision for physical observables. On the other hand, our
result removes any uncertainties on the value of the form factor
present in Ref.~\cite{Davies:2019nhm}. Once b-quark
loops are considered at non-vanishing b-quark mass, our result becomes
indispensable, since Pad\'e approximants potentially induce errors on
physical predictions in the ten-precent range.

For the presentation of our results, we have used two different
infrared-renormalisation schemes. On the other hand, we have chosen to
renormalise the Yukawa coupling in the on-shell scheme. Fortunately, a
translation to any other scheme, e.g.\ $\overline{\mathrm{MS}}$, can
be easily achieved thanks to the knowledge of one- and two-loop results
in analytic form. This translation is independent of infrared
renormalisation.

In principle, our calculation can also be used to obtain the form
factor for the process $H \to \gamma\gamma$, as well as processes
involving pseudo-scalars instead of a scalar. We intend to provide
these results in forthcoming publications.

Finally, we stress that a complete knowledge of the form factor at
three loops in the most general case requires the evaluation of
diagrams with two different massive quarks. This can be achieved with
numerical methods presented here, for example by fixing the ratio of
the b- and top-quark masses. We leave this problem to future work.

Our results are available in computer readable form, see
Appendix~\ref{sec:supplemental}.


\section*{Acknowledgements}

This work was supported by the Deutsche Forschungsgemeinschaft under
grant 396021762 - TRR 257.


\newpage

\appendix


\section{Large-mass expansion} \label{sec:LME}

\be
C_I^{(2,0)} = \sum_{n=0}^\infty \left( a_{n,0} + a_{n,1} \, L_s\right) \, z^n
\; , \qquad
L_s \equiv \ln\left( -\frac{s}{M^2} - i0^+ \right) \; ,
\ee
\begin{align}
\begin{autobreak}
\mathcal{C}^{(2,0)}_I =
10.1151523
+0.3958333333 \, L_s
+\Big(4.778475062
+0.6374228395 \, L_s
\Big) \, z
+\Big(3.071997564
+0.3726469724 \, L_s
\Big) \, z^{2}
+\Big(2.113752253
+0.2432786092 \, L_s
\Big) \, z^{3}
+\Big(1.549293473
+0.1705037577 \, L_s
\Big) \, z^{4}
+\Big(1.188613713
+0.1259718957 \, L_s
\Big) \, z^{5}
+\Big(0.9434907022
+0.09730796586 \, L_s
\Big) \, z^{6}
+\Big(0.7698982981
+0.07720332487 \, L_s\Big) \, z^{7}
+\Big(0.6413834263
+0.06309263508 \, L_s
\Big) \, z^{8}
+\Big(0.5441670543
+0.05233299715 \, L_s
\Big) \, z^{9}
+\Big(0.4680660377
+0.0443515761 \, L_s\Big) \, z^{10}
+\Big(0.4078535762
+0.0379145141 \, L_s
\Big) \, z^{11}
+\Big(0.3588441671
+0.03295598554 \, L_s
\Big) \, z^{12}
+\Big(0.3187910863
+0.02879246954 \, L_s\Big) \, z^{13}
+\Big(0.2852395627
+0.02549720532 \, L_s
\Big) \, z^{14}
+\Big(0.2571429415
+0.02264475648 \, L_s
\Big) \, z^{15}
+\Big(0.2330827097
+0.02034099839 \, L_s\Big) \, z^{16}
+\Big(0.2125481506
+0.01829860319 \, L_s
\Big) \, z^{17}
+\Big(0.1946546291
+0.01662316981 \, L_s
\Big) \, z^{18}
+\Big(0.1791494756
+0.01510884572 \, L_s\Big) \, z^{19}
+\Big(0.165446505
+0.01385124062 \, L_s
\Big) \, z^{20}
+\Big(0.1534242422
+0.01269623978 \, L_s
\Big) \, z^{21}
+\Big(0.1426747173
+0.01172751873 \, L_s\Big) \, z^{22}
+\Big(0.1331457079
+0.0108257358 \, L_s
\Big) \, z^{23}
+\Big(0.1245416017
+0.01006326825 \, L_s
\Big) \, z^{24}
+\Big(0.1168475512
+0.009345212464 \, L_s\Big) \, z^{25}
+\Big(0.1098420694
+0.008734026363 \, L_s
\Big) \, z^{26}
+\Big(0.1035305936
+0.00815260668 \, L_s
\Big) \, z^{27}
+\Big(0.09774245353
+0.007654956458 \, L_s
\Big) \, z^{28}
+\Big(0.09249392018
+0.007177321845 \, L_s
\Big) \, z^{29}
+\Big(0.08765032243
+0.006766580255 \, L_s
\Big) \, z^{30}
+\Big(0.08323342197
+0.006369234271 \, L_s
\Big) \, z^{31}
+\Big(0.07913478115
+0.006026173094 \, L_s
\Big) \, z^{32}
+\Big(0.07537859115
+0.005691941208 \, L_s
\Big) \, z^{33}
+\Big(0.07187600638
+0.005402388748 \, L_s
\Big) \, z^{34}
+\Big(0.068651888
+0.005118475844 \, L_s
\Big) \, z^{35}
+\Big(0.06563233955
+0.004871797509 \, L_s
\Big) \, z^{36}
+\Big(0.06284189053
+0.004628509056 \, L_s
\Big) \, z^{37}
+\Big(0.06021826802
+0.004416595976 \, L_s
\Big) \, z^{38}
+\Big(0.05778511593
+0.004206475477 \, L_s
\Big) \, z^{39}
+\Big(0.0554893509
+0.004023055687 \, L_s
\Big) \, z^{40}
+\Big(0.05335344447
+0.003840290207 \, L_s
\Big) \, z^{41}
+\Big(0.05133168673
+0.003680449807 \, L_s
\Big) \, z^{42}
+\Big(0.04944524976
+0.003520452297 \, L_s
\Big) \, z^{43}
+\Big(0.04765441847
+0.003380296411 \, L_s
\Big) \, z^{44}
+\Big(0.04597903331
+0.003239407083 \, L_s
\Big) \, z^{45}
+\Big(0.04438431107
+0.003115815786 \, L_s
\Big) \, z^{46}
+\Big(0.04288878386
+0.002991085231 \, L_s
\Big) \, z^{47}
+\Big(0.04146176925
+0.002881535172 \, L_s
\Big) \, z^{48}
+\Big(0.04012054547
+0.00277056456 \, L_s
\Big) \, z^{49}
+\Big(0.03883787271
+0.002672996802 \, L_s
\Big) \, z^{50}
+\Big(0.03762984606
+0.002573818605 \, L_s
\Big) \, z^{51}
+\Big(0.03647214107
+0.002486539456 \, L_s
\Big) \, z^{52}
+\Big(0.03537974752
+0.002397527326 \, L_s
\Big) \, z^{53}
+\Big(0.03433082942
+0.002319133078 \, L_s
\Big) \, z^{54}
+\Big(0.03333935067
+0.002238933009 \, L_s
\Big) \, z^{55}
+\Big(0.03238561472
+0.00216825219 \, L_s
\Big) \, z^{56}
+\Big(0.0314826373
+0.002095729392 \, L_s
\Big) \, z^{57}
+\Big(0.03061257345
+0.002031778036 \, L_s
\Big) \, z^{58}
+\Big(0.0297875648
+0.001965975545 \, L_s
\Big) \, z^{59}
+\Big(0.02899137863
+0.001907922209 \, L_s
\Big) \, z^{60}
+\Big(0.028235352
+0.001848028357 \, L_s
\Big) \, z^{61}
+\Big(0.02750466385
+0.001795166544 \, L_s
\Big) \, z^{62}
+\Big(0.02680991123
+0.001740489184 \, L_s
\Big) \, z^{63}
+\Big(0.02613751555
+0.001692215416 \, L_s
\Big) \, z^{64}
+\Big(0.02549739331
+0.001642161338 \, L_s
\Big) \, z^{65}
+\Big(0.02487706468
+0.001597957601 \, L_s
\Big) \, z^{66}
+\Big(0.02428582021
+0.001552015981 \, L_s
\Big) \, z^{67}
+\Big(0.02371215609
+0.00151143557 \, L_s
\Big) \, z^{68}
+\Big(0.02316478676
+0.001469164564 \, L_s
\Big) \, z^{69}
+\Big(0.02263307899
+0.001431820595 \, L_s
\Big) \, z^{70}
+\Big(0.02212521664
+0.001392836423 \, L_s
\Big) \, z^{71}
+\Big(0.02163134597
+0.001358392448 \, L_s
\Big) \, z^{72}
+\Big(0.02115916176
+0.001322360441 \, L_s
\Big) \, z^{73}
+\Big(0.02069951073
+0.001290522743 \, L_s
\Big) \, z^{74}
+\Big(0.02025963637
+0.001257149966 \, L_s
\Big) \, z^{75}
+\Big(0.01983101697
+0.001227661193 \, L_s
\Big) \, z^{76}
+\Big(0.01942047911
+0.001196690333 \, L_s
\Big) \, z^{77}
+\Big(0.01902007236
+0.001169324218 \, L_s
\Big) \, z^{78}
+\Big(0.01863623774
+0.00114052849 \, L_s
\Big) \, z^{79}
+\Big(0.01826154308
+0.00111508544 \, L_s
\Big) \, z^{80}
+\Big(0.01790207242
+0.001088264332 \, L_s
\Big) \, z^{81}
+\Big(0.01755086504
+0.001064567732 \, L_s
\Big) \, z^{82}
+\Big(0.01721367407
+0.001039543422 \, L_s
\Big) \, z^{83}
+\Big(0.01688396883
+0.001017436529 \, L_s
\Big) \, z^{84}
+\Big(0.01656719533
+0.0009940508608 \, L_s
\Big) \, z^{85}
+\Big(0.01625721611
+0.000973394174 \, L_s
\Big) \, z^{86}
+\Big(0.01595919178
+0.000951506098 \, L_s
\Big) \, z^{87}
+\Big(0.01566734531
+0.0009321751265 \, L_s
\Big) \, z^{88}
+\Big(0.0153865718
+0.0009116585178 \, L_s
\Big) \, z^{89}
+\Big(0.01511142531
+0.00089354188 \, L_s
\Big) \, z^{90}
+\Big(0.01484655363
+0.0008742836775 \, L_s
\Big) \, z^{91}
+\Big(0.01458681559
+0.0008572814757 \, L_s
\Big) \, z^{92}
+\Big(0.0143366284
+0.0008391800862 \, L_s
\Big) \, z^{93}
+\Big(0.01409113199
+0.0008232025116 \, L_s
\Big) \, z^{94}
+\Big(0.01385452828
+0.0008061664372 \, L_s
\Big) \, z^{95}
+\Big(0.01362221703
+0.0007911325696 \, L_s
\Big) \, z^{96}
+\Big(0.01339819894
+0.0007750792236 \, L_s
\Big) \, z^{97}
+\Big(0.01317811431
+0.000760915994 \, L_s
\Big) \, z^{98}
+\Big(0.01296577557
+0.0007457706765 \, L_s
\Big) \, z^{99}
+\Big(0.01275704613
+0.0007324119688 \, L_s
\Big) \, z^{100}

+ \order{z^{101}} \; .
\end{autobreak}
\end{align}
The exact expansion coefficients are provided in
Ref.~\cite{ggHCzakonNiggetiedt}. We agree with
Refs.~\cite{Harlander:2009bw, Pak:2009bx} up to $\order{z^4}$ and with
Ref.~\cite{Davies:2019nhm} up to $\order{z^6}$.


\newpage

\section{Threshold expansion} \label{sec:THR}

\begin{figure}[t]
\center
\includegraphics[width=\textwidth]{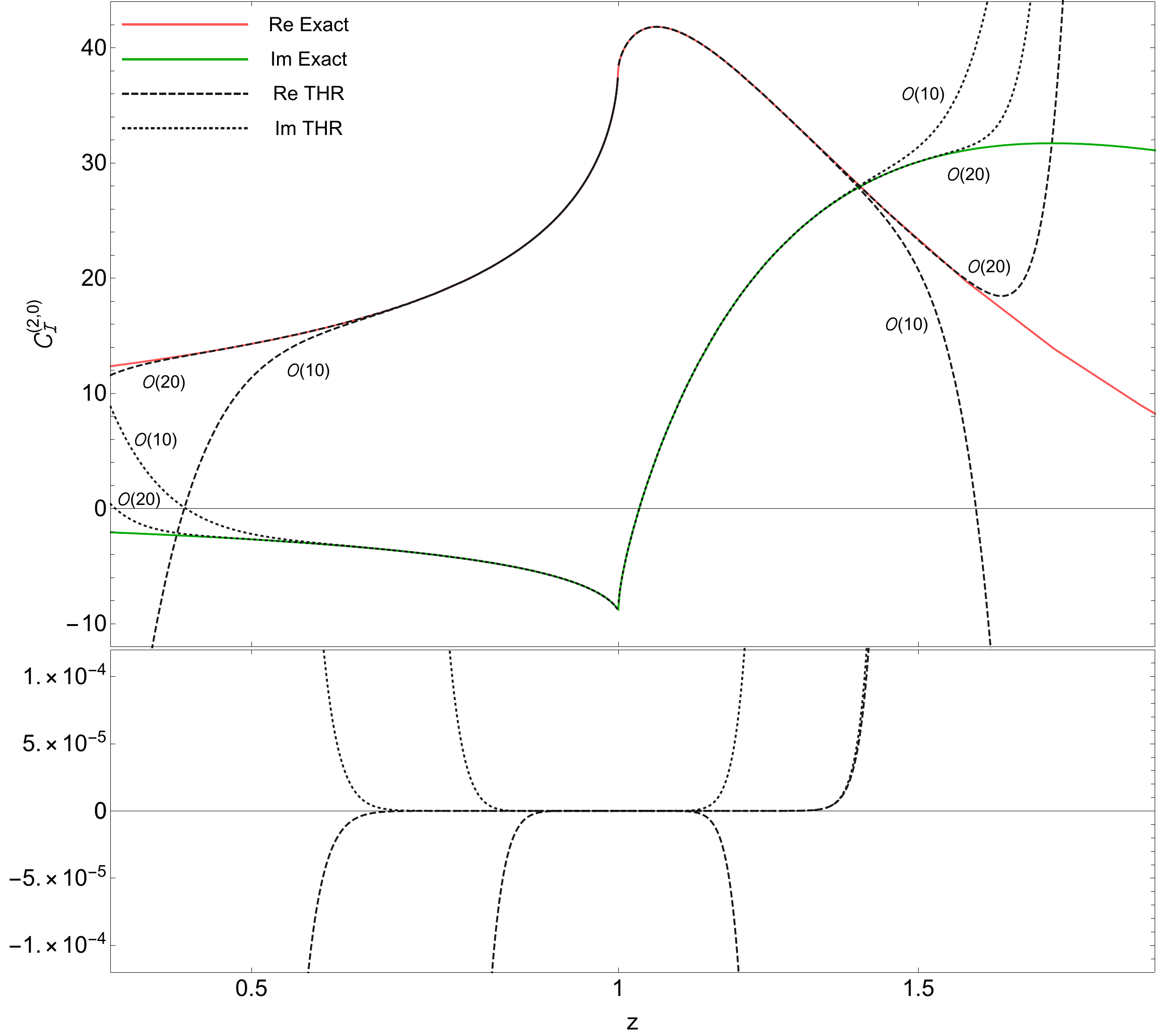}
\caption{Comparison of the threshold expansion (THR) truncated at
  $\order{(1-z)^{10}}$ and $\order{(1-z)^{20}}$ with the exact result
  for the three-loop coefficient of the finite remainder
  $\mathcal{C}_I^{(2)}$ at $n_l = 0$, $L_\mu = 0$. The lower panel
  shows the absolute difference between the expansions and the exact
  result.}
\label{fig:THR}
\end{figure}
\be
\begin{aligned}
&C_I^{(2,0)} = \sum_{n=0}^\infty \left( b_{n,0} + b_{n,1} \, L_t +
  b_{n,2} \, L_t^2 \right) \, t^n \; , \\[.2cm]
&L_t \equiv \ln\left( 1 - z \right) \; , \qquad t \equiv
\sqrt{1 - z} = \exp(L_t/2) \; ,
\end{aligned}
\ee
\begin{align}
\begin{autobreak}
\mathcal{C}^{(2,0)}_I =
38.29655119
-8.9070147 \, i
-29.55840851 \, t
+\Big(9.112936321
-68.1395365 \, i
+(14.16269653
-28.42242029 \, i) \, L_t
-4.523568684 \, L_t^2
\Big) \, t^{2}
+\Big(-20.55378026
+133.7985485 \, i
-26.60436928 \, L_t
\Big) \, t^{3}
+\Big(-25.39554578
-239.3964484 \, i
+(14.71881407
-18.94828019 \, i) \, L_t
-8.864366916 \, L_t^2
\Big) \, t^{4}
+\Big(22.88555562
+311.994478 \, i
+(-43.65929113
-30.41485955 \, i) \, L_t
+(-0.3490658504
+7.402203301 \, i) \, L_t^2
\Big) \, t^{5}
+\Big(-122.1397994
-392.2909322 \, i
+(6.009726459
+13.26379614 \, i) \, L_t
-5.516621472 \, L_t^2
\Big) \, t^{6}
+\Big(140.6543286
+457.2900946 \, i
+(-70.34961079
-68.49797789 \, i) \, L_t
+(2.520477069
+19.98594891 \, i) \, L_t^2
\Big) \, t^{7}
+\Big(-310.5867852
-492.0494746 \, i
+(-6.024184876
+62.80001436 \, i) \, L_t
+7.272523314 \, L_t^2
\Big) \, t^{8}
+\Big(359.8673214
+541.6828656 \, i
+(-116.1605627
-105.7705087 \, i) \, L_t
+(10.80021746
+36.50872414 \, i) \, L_t^2
\Big) \, t^{9}
+\Big(-610.5588771
-520.6092531 \, i
+(-16.89694249
+126.7730175 \, i) \, L_t
+30.31073877 \, L_t^2
\Big) \, t^{10}
+\Big(700.264643
+549.2184102 \, i
+(-185.5662205
-138.0922834 \, i) \, L_t
+(26.001469
+56.2053409 \, i) \, L_t^2
\Big) \, t^{11}
+\Big(-1036.677064
-465.6193352 \, i
+(-22.72070802
+203.3453965 \, i) \, L_t
+64.09089853 \, L_t^2
\Big) \, t^{12}
+\Big(1177.227509
+468.516149 \, i
+(-279.6079413
-163.1047377 \, i) \, L_t
+(49.23219084
+78.56007164 \, i) \, L_t^2
\Big) \, t^{13}
+\Big(-1600.347449
-317.3641993 \, i
+(-19.90826165
+291.2249078 \, i) \, L_t
+108.9603809 \, L_t^2
\Big) \, t^{14}
+\Big(1803.642376
+290.4177751 \, i
+(-396.3597986
-179.328259 \, i) \, L_t
+(81.34424135
+103.2021919 \, i) \, L_t^2
\Big) \, t^{15}
+\Big(-2310.834851
-67.58556999 \, i
+(-5.061769755
+389.4423686 \, i) \, L_t
+165.1852917 \, L_t^2
\Big) \, t^{16}
+\Big(2591.104007
+7.032756569 \, i
+(-530.7745415
-185.7721012 \, i) \, L_t
+(123.0180749
+129.8524061 \, i) \, L_t^2
\Big) \, t^{17}
+\Big(-3175.897771
+291.0629642 \, i
+(25.07572246
+497.2376355 \, i) \, L_t
+232.9804921 \, L_t^2
\Big) \, t^{18}
+\Big(3550.803491
-388.7191604 \, i
+(-674.0712248
-181.7412648 \, i) \, L_t
+(174.8140502
+158.2927098 \, i) \, L_t^2
\Big) \, t^{19}
+\Big(-4202.187543
+765.3035663 \, i
+(73.64516718
+613.9945923 \, i) \, L_t
+312.5257534 \, L_t^2
\Big) \, t^{20}
+\Big(4694.305121
-903.3548866 \, i
+(-812.6741315
-166.7311215 \, i) \, L_t
+(237.2049629
+188.348174 \, i) \, L_t^2
\Big) \, t^{21}
+\Big(-5395.511981
+1361.401213 \, i
+(143.6864402
+739.2011425 \, i) \, L_t
+403.9753603 \, L_t^2
\Big) \, t^{22}
+\Big(6034.345387
-1542.973831 \, i
+(-926.6236752
-140.3655593 \, i) \, L_t
+(310.5975616
+219.8752962 \, i) \, L_t^2
\Big) \, t^{23}
+\Big(-6761.017023
+2085.268299 \, i
+(238.1485965
+872.4231173 \, i) \, L_t
+507.4642605 \, L_t^2
\Big) \, t^{24}
+\Big(7585.760424
-2313.350376 \, i
+(-987.3075061
-102.3585515 \, i) \, L_t
+(395.3473322
+252.7542072 \, i) \, L_t^2
\Big) \, t^{25}
+\Big(-8303.316472
+2942.534197 \, i
+(359.897675
+1013.286463 \, i) \, L_t
+623.1122305 \, L_t^2
\Big) \, t^{26}
+\Big(9366.647246
-3219.995987 \, i
+(-954.2825083
-52.48916829 \, i) \, L_t
+(491.7690072
+286.8832593 \, i) \, L_t^2
\Big) \, t^{27}
+\Big(-10026.58756
+3938.593854 \, i
+(511.7230633
+1161.464627 \, i) \, L_t
+751.026828 \, L_t^2
\Big) \, t^{28}
+\Big(11399.88074
-4268.202962 \, i
+(-770.8622683
+9.415287382 \, i) \, L_t
+(600.1442562
+322.175152 \, i) \, L_t^2
\Big) \, t^{29}
+\Big(-11934.64323
+5078.643159 \, i
+(696.3429187
+1316.669338 \, i) \, L_t
+891.3055595 \, L_t^2
\Big) \, t^{30}
+\Big(13715.1406
-5463.076486 \, i
+(-358.0179892
+83.49101847 \, i) \, L_t
+(720.7274605
+358.554085 \, i) \, L_t^2
\Big) \, t^{31}
+\Big(-14030.98797
+6367.70559 \, i
+(916.4089125
+1478.643714 \, i) \, L_t
+1044.037518 \, L_t^2
\Big) \, t^{32}
+\Big(16351.65146
-6809.558924 \, i
+(394.0297813
+169.8461219 \, i) \, L_t
+(853.750144
+395.953618 \, i) \, L_t^2
\Big) \, t^{33}
+\Big(-16318.86196
+7810.652869 \, i
+(1174.510449
+1647.156969 \, i) \, L_t
+1209.304654 \, L_t^2
\Big) \, t^{34}
+\Big(19361.90922
-8312.448766 \, i
+(1637.383024
+268.5667181 \, i) \, L_t
+(999.4244394
+434.3150324 \, i) \, L_t^2
\Big) \, t^{35}
+\Big(-18801.27633
+9412.2214 \, i
+(1473.178459
+1822.0003 \, i) \, L_t
+1387.182774 \, L_t^2
\Big) \, t^{36}
+\Big(22816.76328
-9976.415748 \, i
+(3582.371933
+379.721518 \, i) \, L_t
+(1157.945845
+473.5860544 \, i) \, L_t^2
\Big) \, t^{37}
+\Big(-21481.04193
+11177.0256 \, i
+(1814.888837
+2002.983613 \, i) \, L_t
+1577.742354 \, L_t^2
\Big) \, t^{38}
+\Big(26812.35732
-11806.01319 \, i
+(6519.39523
+503.3652836 \, i) \, L_t
+(1329.495449
+513.7198453 \, i) \, L_t^2
\Big) \, t^{39}
+\Big(-24360.79309
+13109.56893 \, i
+(2202.065568
+2189.932902 \, i) \, L_t
+1781.049193 \, L_t^2
\Big) \, t^{40}

+ \order{t^{41}} \; .
\end{autobreak}
\end{align}
We agree with Ref.~\cite{Grober:2017uho} for the coefficients of the
first three non-analytic terms:
\be
b_{1,0} = -\frac{2\pi^3}{27}(3+\pi^2) \; , \quad
b_{2,1} = \frac{\pi^2}{216}(458-15\pi^2) +2 \pi i \, b_{2,2} \qand
b_{2,2} = -\frac{99\pi^2}{216} \; .
\ee
We also provide a high precision result for the three-loop coefficient
of the form-factor at threshold:
\be
\begin{split}
\mathcal{C}_I^{(2,0)}\big[ z = 1 \big] = b_{0,0} &\approx
+38.29655118857344308946576090253939 \\[.2cm]
&\quad -8.907014700051001636660098822811295 \, i
\; .
\end{split}
\ee
%


\newpage

\section{High-energy expansion} \label{sec:HE}

\begin{figure}[t]
\center
\includegraphics[width=\textwidth]{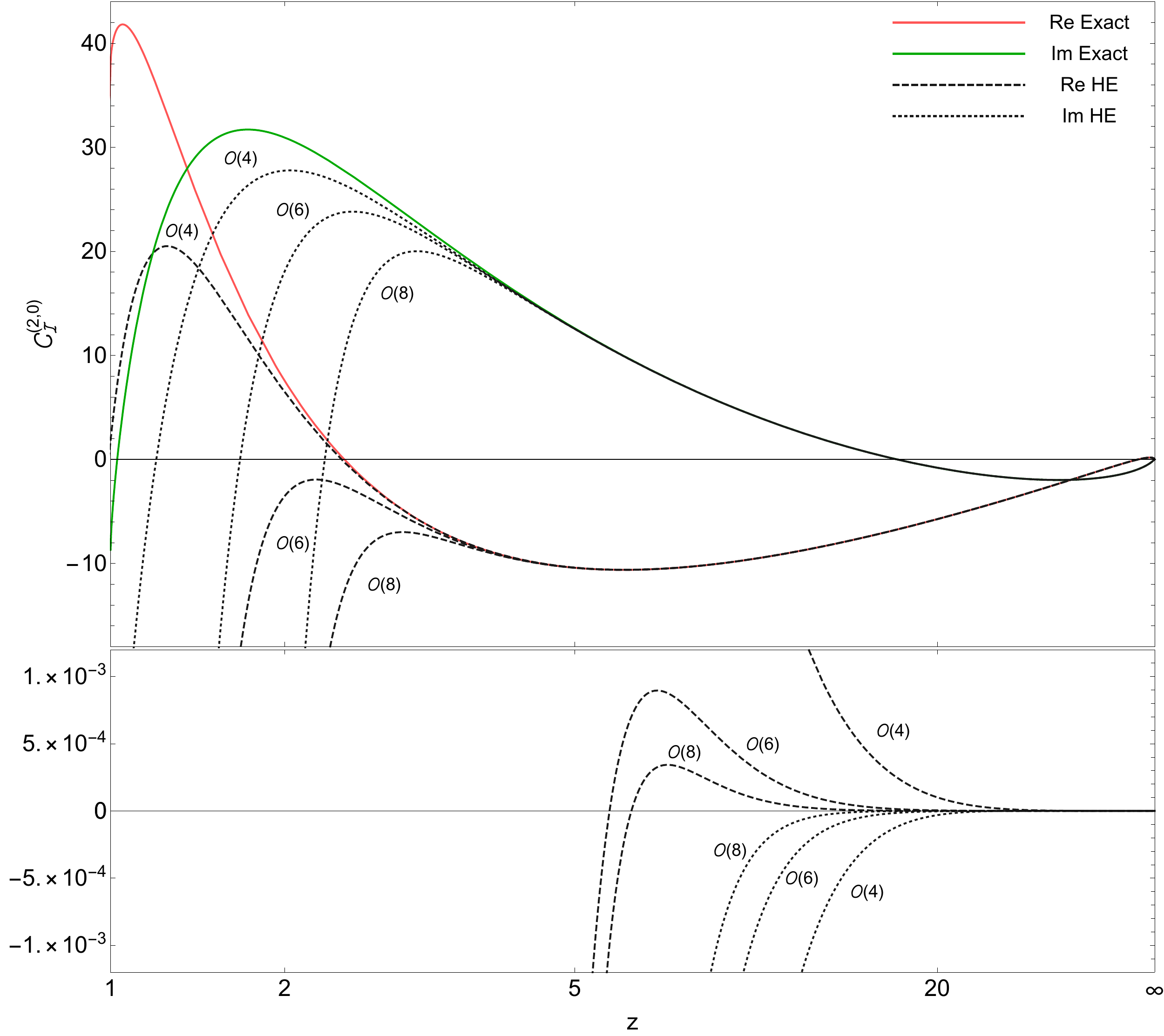}
\caption{Comparison of the high-energy expansion (HE) truncated at
  $\order{1/z^{4}}$, $\order{1/z^{6}}$ and $\order{1/z^{8}}$ with the
  exact result for the three-loop coefficient of the finite remainder
  $\mathcal{C}_I^{(2)}$ at $n_l = 0$, $L_\mu = 0$. The lower panel
  shows the absolute difference between the expansions and the exact
  result.}
\label{fig:HE}
\end{figure}
\be
C_I^{(2,0)} = \sum_{n=1}^\infty \sum_{k=0}^6 c_{n,k} \, L^k_s \, z^{-n}
\; , \qquad
L_s \equiv \ln\left( -\frac{s}{M^2} - i0^+ \right) \; ,
\ee
\begin{align}
\begin{autobreak}
\mathcal{C}^{(2,0)}_I =
\Big(15.93205751
-15.73631507 \, L_s
-1.121722806 \, L_s^2
+0.4035518803 \, L_s^3
+0.08901988687 \, L_s^4
-0.001736111111 \, L_s^5
-0.0004822530864 \, L_s^6
\Big) \, z^{-1}
+\Big(0.06309685356
+3.546786436 \, L_s
-0.519984143 \, L_s^2
-1.652739942 \, L_s^3
-0.1240600623 \, L_s^4
-0.004134114583 \, L_s^5
+0.0005738811728 \, L_s^6
\Big) \, z^{-2}
+\Big(5.754168857
+7.325854683 \, L_s
-2.98120415 \, L_s^2
+0.1651932919 \, L_s^3
+0.003161112205 \, L_s^4
-0.005756293403 \, L_s^5
+0.000220630787 \, L_s^6
\Big) \, z^{-3}
+\Big(-10.66566232
-10.56571524 \, L_s
+10.33923567 \, L_s^2
-0.313124275 \, L_s^3
-0.1681889443 \, L_s^4
+0.01392927758 \, L_s^5
+0.0000316478588 \, L_s^6
\Big) \, z^{-4}
+\Big(-6.785278289
+88.43750151 \, L_s
-40.26616919 \, L_s^2
+2.072111298 \, L_s^3
+0.7341214981 \, L_s^4
-0.04301260489 \, L_s^5
-0.0003223560475 \, L_s^6
\Big) \, z^{-5}
+\Big(80.70142226
-421.2250932 \, L_s
+175.4294283 \, L_s^2
-5.805171716 \, L_s^3
-3.062956746 \, L_s^4
+0.1753725462 \, L_s^5
+0.0009707792306 \, L_s^6
\Big) \, z^{-6}
+\Big(-486.1362845
+2151.385984 \, L_s
-853.4135303 \, L_s^2
+26.4094276 \, L_s^3
+14.84667539 \, L_s^4
-0.8733492022 \, L_s^5
-0.002646085951 \, L_s^6
\Big) \, z^{-7}
+\Big(2880.610148-11795.75065 \, L_s
+4569.562554 \, L_s^2
-140.0597361 \, L_s^3
-78.99328343 \, L_s^4
+4.758979333 \, L_s^5
+0.008394276654 \, L_s^6
\Big) \, z^{-8}

+ \order{z^{-9}} \; .
\end{autobreak}
\end{align}
The value of the coefficient of the term proportional to $L_s^6/z$
agrees with Refs.~\cite{Liu:2017vkm, Liu:2018czl}, while the
coefficient of the term proportional to $L_s^5/z$ has been confirmed in
Ref.~\cite{Anastasiou:2020vkr}.


\section{Supplemental material} \label{sec:supplemental}

The ancillary file, Ref.~\cite{ggHCzakonNiggetiedt}, conforming to
\textsc{Wolfram Mathematica} format, provides the following results as
second order polynomials in \verb|api|$\equiv \alpha_s/\pi$:
\begin{description}
\item \verb|CI[z, nl, Lmu]| - $\mathcal{C}_I$, Eq.~\eqref{eq:finiteRemainderI};
\item \verb|CZ[z, nl, Lmu]| - $\mathcal{C}_Z$, Eq.~\eqref{eq:finiteRemainderZ};
\item \verb|CItoCZ| - conversion between infrared schemes,
  Eq.~\eqref{eq:conversion}.
\end{description}
The approximations used by the function \verb|CI[z, nl, Lmu]| are
directly accessible with the following functions evaluated at $L_\mu =0$:
\begin{description}
\item \verb|C0[z], C1I[z], C2I[z, nl]| - $\mathcal{C}^{(0)}$,
  $\mathcal{C}_I^{(1)}$ and $\mathcal{C}_I^{(2)}$,
  Eqs.~\eqref{eq:expansion}~and~\eqref{eq:finiteRemainderI};
\item \verb|C2ILMEnl0[z], C2ILMEnl1[z]| - large-mass expansion of
  $\mathcal{C}_I^{(2,0)}$ (Appendix~\ref{sec:LME}) and
  $\mathcal{C}_I^{(2,1)}$;
\item \verb|C2ITHRnl0[z], C2ITHRnl1[z]| - threshold expansion of
  $\mathcal{C}_I^{(2,0)}$ (Appendix~\ref{sec:THR}) and
  $\mathcal{C}_I^{(2,1)}$;
\item \verb|C2IHEnl0[z], C2IHEnl1[z]| - high-energy expansion of
  $\mathcal{C}_I^{(2,0)}$ (Appendix~\ref{sec:HE}) and
  $\mathcal{C}_I^{(2,1)}$;
\item \verb|C2ITABnl0[z], C2ITABnl1[z]| - interpolation of
  $\mathcal{C}_I^{(2,0)}$ (Tabs.~\ref{tab:num1}~and~\ref{tab:num2})
  and $\mathcal{C}_I^{(2,1)}$.
\end{description}
All functions require a numeric value for $z$. Finally, the large-mass
expansion of $\mathcal{C}_I^{(2)}$ evaluated at $L_\mu =0$ with exact
coefficients and dependence on $n_l$ is given by \verb|C2ILME|.

\noindent
The results correspond to QCD with $C_A = 3$, $C_F = 4/3$, $T_F = 1/2$.

\noindent
Note that we do not use the results of Ref.~\cite{Harlander:2019ioe}
for $\mathcal{C}_I^{(2,1)}$ in the ancillary file.


\newpage

\bibliographystyle{JHEP}
\bibliography{ggH}

\end{document}